\documentclass[acmsmall]{acmart}
\usepackage{graphicx}
\usepackage{algorithm}
\usepackage{listings}
\usepackage{gensymb}
\usepackage[utf8]{inputenc}
\DeclareUnicodeCharacter{2212}{-}
\usepackage{mathtools}
\usepackage{algpseudocode}
\usepackage{multirow}
\usepackage{caption}
\usepackage{subcaption}
\usepackage{svg}
\usepackage[flushleft]{threeparttable}
\usepackage[T1]{fontenc}
\usepackage{titlesec}
\titleformat{\paragraph}[runin]
{\bfseries}{\theparagraph}{1em}{}

\AtBeginDocument{%
  \providecommand\BibTeX{{%
    \normalfont B\kern-0.5em{\scshape i\kern-0.25em b}\kern-0.8em\TeX}}}

\setcopyright{none}
\acmYear{2022}
\acmDOI{XXXXXXX.XXXXXXX}

\acmJournal{TECS}

\begin{document}

\title{An End-to-End Analysis of EMFI on Bit-sliced Post-Quantum Implementations}

\author{Richa Singh}
\email{rsingh7@wpi.edu}
\affiliation{%
  \institution{Worcester Polytechnic Institute}
  \streetaddress{100, Institute Road}
  \city{Worcester}
  \state{Massachusetts}
  \country{USA}
  \postcode{01609}
}

\author{Saad Islam}
\email{sislam@wpi.edu}
\affiliation{%
  \institution{Worcester Polytechnic Institute}
  \streetaddress{100, Institute Road}
  \city{Worcester}
  \state{Massachusetts}
  \country{USA}
  \postcode{01609}
}

\author{Berk Sunar}
\email{sunar@wpi.edu}
\affiliation{%
  \institution{Worcester Polytechnic Institute}
  \streetaddress{100, Institute Road}
  \city{Worcester}
  \state{Massachusetts}
  \country{USA}
  \postcode{01609}
}

\author{Patrick Schaumont}
\email{pschaumont@wpi.edu}
\affiliation{%
  \institution{Worcester Polytechnic Institute}
  \streetaddress{100, Institute Road}
  \city{Worcester}
  \state{Massachusetts}
  \country{USA}
  \postcode{01609}
}

\renewcommand{\shortauthors}{Richa, et al.}

\begin{abstract}
Bit-slicing is a software implementation technique that treats an N-bit processor datapath as N parallel single-bit datapaths. The natural spatial redundancy of bit-sliced software can be used to build countermeasures against implementation attacks. While the merits of bit-slicing for side-channel countermeasures have been studied before, their application for protection of post-quantum algorithms against fault injection is still unexplored. We present an end-to-end analysis of the efficacy of bit-slicing to detect and thwart electromagnetic fault injection (EMFI) attacks on post-quantum cryptography (PQC). We study Dilithium, a digital signature finalist of the NIST PQC competition. We present a bit-slice-redundant design for the Number-Theoretic Transform (NTT), the most complex and compute-intensive component in Dilithium. We show a data-redundant countermeasure for NTT which offers two concurrent bits for every single bit in the original implementation. We then implement a full Dilithium signature sequence on a 667 MHz ARM Cortex-A9 processor integrated in a Xilinx Zynq SoC. We perform a detailed EM fault-injection parameter search to optimize the location, intensity and timing of injected EM pulses. We demonstrate that, under optimized fault injection parameters, about 10\% of the injected faults become potentially exploitable. However, the bit-sliced NTT design is able to catch the majority of these potentially exploitable faults, even when the remainder of the Dilithium algorithm as well as the control flow is left unprotected. To our knowledge, this is the first demonstration of a bitslice-redundant design of Dilithium that offers distributed fault detection throughout the execution of the algorithm.

%
\end{abstract}

\ccsdesc[500]{Security and Privacy~Post-Quantum Lattice-Based Cryptography}
\ccsdesc[300]{Security and Privacy~Digital Signatures}
\ccsdesc{Security and Privacy~Fault Attacks and Countermeasures}

\keywords{Dilithium, Bit-slicing, Intra-Instruction Redundancy, Fault Attack Countermeasure, ARM Cortex-A9, Electromagnetic Fault Injection}

\maketitle

\section{Introduction}

Due to advancements in Quantum Computing and the discovery of Shor's algorithm \cite{10.1007/3-540-58691-1_68}, there is a looming threat to our existing public key infrastructure (PKI). With the possibility of powerful and large-scale quantum computers in not too distant future, common RSA- and ECC-based protocols become vulnerable.

NIST recently initiated a Post-Quantum Cryptography (PQC) Standardization process to select and standardize quantum-resistant algorithms \cite{257451} for Public Key Encryption (PKE), Key Establishment Mechanisms (KEM) and Digital Signatures (DS) in order to replace RSA and ECC schemes. This process is currently in its third and final round which has three finalist candidates in DS scheme category: Dilithium, Falcon and Rainbow \cite{257451}. The criterion that determines the selection process includes classical as well as post-quantum (PQ) security guarantees.
For the final round, in addition to key selection criterion such as classical security, theoretical PQ security guarantees, implementation cost and performance; resistance against active and passive implementation attacks, one of the most important criteria \cite{257451}. Several authors  have reported physical attacks on structured lattice-based schemes through exploitation of a number of side-channels such as power \cite{DBLP:journals/iacr/Pessl17, 10.1007/978-3-319-66787-4_25, 10.1007/978-3-030-30530-7_7, ravi2018side, 8383894, mag_sca, 9643637}, electromagnetic emissions \cite{espitau:hal-01648080}, cache timing \cite{blisscacheattack, 10.1145/3133956.3134023} and induced faults \cite{bruinderink2018differential, tubiblio104111, asiacrypt-2021-31371, 10.1145/3178291.3178294, 10.1007/978-3-319-69453-5_8, ravi2019number, DBLP:journals/tches/PesslP21, ravi2018side, ravi2019exploiting}.

In real-world usage scenarios, DS schemes will be deployed on a wide variety of platforms, including personal computers (PCs) and resource-constrained devices. It is crucial to analyze whether DS primitives and their implementations are vulnerable against fault injection analysis. There are previous works that have analyzed fault vulnerabilities on deterministic variants of lattice-based schemes to recover a part of secret key and validated those fault vulnerabilities using electromagnetic fault injection (EMFI) \cite{ravi2019number, ravi2019exploiting} and clock fault injection \cite{bruinderink2018differential, asiacrypt-2021-31371} attacks. Such attacks are not impeded by basic fault countermeasures such as assembly-level instruction duplication/triplication \cite{10.1145/1873548.1873555}, algorithm-level temporal redundancy \cite{symkey_cm, Ciet2005PracticalFC, faultcm_guide} and verification-after-sign \cite{ducas2018crystals}. A major problem with these time-based redundancy techniques is that due to modern fault-injection capabilities, an adversary can inject consistent faults in redundant sections of code or data that may lead to faults go undetected. An adversary with precise control over the timing of the fault injection could bypass verification-over-sign countermeasure when fault effects are modelled by an instruction-skip. 

The Number Theoretic Transform (NTT) is a central computation step in many post-quantum schemes including Dilithium. It is used to efficiently perform polynomial multiplication. Recent work has demonstrated that NTT can be used as a side-channel target leading to full-key recovery based on a single trace \cite{10.1007/978-3-319-66787-4_25, 10.1007/978-3-030-30530-7_7}. To defend against such single-trace attacks on NTT, Ravi {\em et al.} proposed masking and shuffling countermeasures \cite{configurable_sca_cm}. The performance overhead of their countermeasure for Dilithium's key generation and signature generation procedures are in the range 12-197\% and 32-490\% respectively. However, this scheme has not been designed with controlled fault injection in mind.

Clearly, there is a need for comprehensive and fine-grained countermeasures to simultaneously protect lattice-based schemes against fault-injection and side-channel attacks while incurring tolerable performance costs. Bit-slicing has been proposed as a concrete software countermeasure against combined fault attacks and side-channel attacks using Intra-Instruction Redundancy (IIR) \cite{DBLP:conf/sacrypt/PatrickYGS16, 10.1007/978-3-030-68773-1_11, 8634950} and higher-order masking \cite{10.1007/978-3-030-68773-1_11} respectively. 

As a first step, we present a bit-sliced design for the most critical sub-block in lattice-based schemes, NTT/INTT. We do so by representing the FFT-like structure of the NTT as parallel butterfly operations in each stage of NTT and then iterating over all the stages to compute final NTT output. We further exploit this bit-sliced design to provide a \textit{spatial intra-instruction redundant} NTT design to protect it against faults in the \textit{datapath}. We incorporate our countermeasure over the unprotected reference implementation of Dilithium Round 3 \footnote{https://github.com/pq-crystals/dilithium}. There is no prior work of fault attacks on the non-deterministic version of Dilithium (Round 3). To experimentally evaluate the effectiveness of the proposed countermeasure, we perform EMFI attack on an ARM Cortex-A9 processor-based Zynq SoC running Dilithium protected with our countermeasure.

\subsection{Our Contribution}
\begin{enumerate}
\item We propose a software countermeasure based on dual data-redundancy obtained through our novel bit-sliced design for NTT in order to detect computation faults. While the use of bit-slicing with Intra-Instruction Redundancy (IIR) in context of fault attacks is not new \cite{10.1007/978-3-030-68773-1_11}, it's application in the context of lattice-based post-quantum algorithms is still unexplored. 
\item The proposed countermeasure is generic and can easily be ported to other lattice-based post-quantum schemes having polynomial multiplications in abundance.
\item We analyze the security of the proposed countermeasure to formulate estimated fault coverage in Dilithium. 
\item We evaluate the effectiveness of the proposed countermeasure on ARMv7-a architecture by testing it against EMFI attack on an 667 MHz ARM-based SoC.
\item We perform a systematic search of the EM fault-injection parameters space, including precise location on the target chip, and optimal ranges for EMFI Pulse Delay and Pulse Power. We are able to increase the percentage of potentially exploitable faults to 10\% of the fault injection attempts. From those potentially exploitable faults, we also determine that about 62\% of them directly affect the data flow. This is possible due to the data-redundant design of our countermeasure to detect all data faults. 


\end{enumerate}

\subsection{Outline}
In Section \ref{section:background}, we will introduce the related literature and relevant background knowledge. Section \ref{section:nttdesign} presents our bit-sliced design for NTT/Inverse NTT and our fault attack countermeasure design based on dual data-redundant bit-slicing. We state our assumptions about attacker's capabilities and fault model in Section \ref{section:fault_attacker_model}. Security analysis of the proposed countermeasure is provided in Section \ref{section:security_analysis}. Section \ref{section:results} provides performance, footprint, overhead and a top-down EMFI analysis of our proposed countermeasure on an ARM-based SoC. Finally, we conclude this paper in Section \ref{section:conclusion}.

\section{Background}
\label{section:background}

This section summarizes important background knowledge concepts, including bit-slicing, intra-instruction redundancy, Electromagnetic fault injection, and the NTT.

\subsection{Bit-slicing}
Bit-slicing is a software optimization technique originally developed to produce high throughput software implementations \cite{DBLP:conf/fse/Biham97a}. It was since also been applied as a side-channel and fault attack countermeasure \cite{DBLP:conf/eurocrypt/BelaidDMRW20,DBLP:conf/sacrypt/PatrickYGS16}.  Bit-sliced code is created from a boolean (bit-level) program.
Bit-sliced programs are written using bit-wise logical operations (AND, OR, NOT, XOR) and as a result they run with bit-level parallelism and can be viewed as follows. An $n$-bit variable in bit-sliced code with bits $b_{{n}-1}...b_1b_0$ is spread over $n$ registers $R_{{n}−1}...R_1R_0$, such that register $R_i$ holds bit $b_i$. A bit-sliced program on an $N$−bit processor computes $N$ instances of $n$-bit variable in parallel because it is composed of $N$ parallel copies of the boolean program. This $N$-fold parallelism can be exploited in NTT computations since each stage of a $2N$-point NTT consists of $N$ independent butterfly operations. NTT operation is mapped to bit-sliced code by computing $N$ instances of butterfly operation per stage in parallel. Applying this bit-sliced butterfly computation over total $\log(2^N)$ stages will result in a $2N$-point NTT output.

\subsection{Bit-slicing with Intra-Instruction Redundancy}
Patrick {\em et al.} proposed a generic fault-attack countermeasure which requires an algorithm to be bit-sliced. We define slice as the bit-location in all $n$ registers which together constitute a $n$-bit variable in bit-sliced code. The parallel bit-slices are used to create redundant copies of original data, which creates data redundancy within the execution of a single bit-wise instruction \cite{DBLP:conf/sacrypt/PatrickYGS16}. The resulting {\em intra-instruction redundancy} thus covers data bits within the same instruction.
To detect faults, the original data slices and its respective redundant slices perform computation on same inputs and at the end of computation, they produce same outputs if there is no fault. 

Intra-instruction redundancy builds on the fact that typical fault injection mechanisms offer poor control over the fault injection location. Therefore, it is difficult to affect the redundant copies within a single word in the same manner. Patrick also discusses an extension of intra-instruction redundancy which protects the control flow of an implementation, ie. the sequence and semantics of bit-wise instructions. In this approach, the different bit-slices allocate time-redundant copies of data bits, such a redundant copy of a data bit belonging to different iterations of an algorithm. 

\subsection{Fault Injection Mechanisms}
Active Fault Attacks (FA) are a well-known threat to embedded devices. In this type of attack, the attacker tries to inject faults in the normal computation by pressuring the underlying device out of its nominal operating conditions. Faulty ciphertext, or fault signatures, can then be used to improve the cryptanalysis. 

Fault injection is a relevant threat to post-quantum cryptography. The deterministic signature of Dilithium was shown susceptible to a Differential Fault Analysis (DFA) \cite{bruinderink2018differential}, and recently also the randomized signature of Dilithium was shown vulnerable to a Signature Correction Attack \cite{DBLP:journals/corr/abs-2203-00637}. 

We discuss some related work relevant to the EM fault injection mechanism used in our work. EMFI attacks produce a local transient magnetic field near the device which induces transient faults into the device computation. This could leak sensitive device information, cause data corruption \cite{8844475}, change the program flow \cite{7140238}, bypass security mechanisms such as secure boot \cite{206182} or cause privileged escalation \cite{9360902}. Our research specifically considers EMFI on ARM architectures. Early work on EMFI is presented in \cite{6305224, Schmidt_opticaland}. Moro {\em et al.} introduced for the first time a precise fault model to study the impact of an EM fault injection on a 32-bit ARM micro-controller \cite{6623558}. Elmohr {\em et al.} shows that EMFI is more susceptible at lower supply voltages and higher clock frequencies on ARM and RISC-V platforms and multi-instruction skips are possible with a single EM pulse \cite{9137051}. Rivi\`ere {\em et al.} performed EM fault injection on cache of an ARM Cortex-M4 micro-controller \cite{7140238}. Menu {\em et al.} presents an EMFI attack on the data-prefetch mechanism of an ARM micro-controller to achieve AES key recovery and AES key resetting with a single fault injection \cite{8844475}. Majeric {\em et al.} successfully injected EM-induced faults into an ARM Cortex-A9 based SoC targeting an AES hardware accelerator \cite{7841183}.

\subsection{Definition of NTT and Inverse NTT}Dilithium uses NTT for efficient polynomial multiplication which is the key part of the scheme and the focus of this work. In this section, we provide the definition of NTT and inverse NTT. Detailed explanation of the structure of Dilithium algorithm can be found in the original specification \cite{ducas2018crystals}.

NTT is a generalized version of well-known Fast-Fourier Transform (FFT) where all the arithmetic is performed in the prime finite field $\mathbb{F}_q$ instead of complex numbers. In ideal lattice-based cryptography, main operation is polynomial multiplication in the ring $R_q = \mathbb{Z}_q[x]/(x^n + 1)$, where, ring $\mathbb{Z}_q[x] = (\mathbb{Z}/q\mathbb{Z})[x]$ denotes the set of polynomials with integer coefficients modulo $q$ and $R_q$ denotes the ring of polynomials with integer coefficients from the ring $\mathbb{Z}_q$ reduced by a $n^{th}$ cyclotomic polynomial $x^n + 1$. NTT-based polynomial multiplication requires that $n$ is a power of two and the modulus $q$ is chosen to be a prime such that $q \equiv 1 \, \text{mod} \, 2n$. This way $\mathbb{Z}_q$ contains a primitive $n$-th root of unity $\omega$ and its square root $\psi$, which means that ${\omega}^n \equiv 1 \, \text{mod} \, q$. 

Let two polynomials $a(x), b(x) \in R_q$ with coefficients $a(x) = (a[0], a[1],\dots, a[n-1])$ and $b(x) = (b[0], b[1],\dots, b[n-1])$ respectively. Then, polynomial multiplication $c = a \cdot b \in R_q$ is defined as 
\begin{equation} 
\label{eq:Polymult_formula}
    c = \text{INTT}(\text{NTT}(a) \odot \text{NTT}(b))
\end{equation}
where, $\odot$ denotes the point-wise multiplication of polynomials.
To eliminate the overhead of zero-padding $a$ and $b$ to length $2n$, negative-wrapped convolution property of NTT is used. So, the transformation $\text{NTT}(a)$ generates a polynomial $\hat{a}$ whose coefficients can be defined as  
\begin{equation}
\label{eq:NTT_formula}
    \hat{a}[i] = \sum_{j=0}^{n-1} {\omega_{n}^{ij}} {\psi^j} {a[j]} \, \text{mod} \, q \; \; \; \forall i \in [0,n-1]
\end{equation}
The inverse NTT is computed in exactly same way except that $\omega_{n}^{ij}$ is replaced with $\omega_{n}^{-ij}$ and the final results are scaled by $\psi^{-i} n^{-1}$ factor. The inverse NTT transforms the polynomial $\hat{a}$ back to $a$ whose coefficients are obtained as
\begin{equation} 
\label{eq:INTT_formula}
    a[i] = {\psi^{-i}} n^{-1} \sum_{j=0}^{n-1} {\omega_{n}^{-ij}} {\hat{a}[j]} \, \text{mod} \, q \; \; \; \forall i \in [0,n-1]
\end{equation}

\section{Proposed Bit-sliced design of NTT-based Polynomial Multiplication}
\label{section:nttdesign}
We propose to use bit-slicing technique on NTT and then utilize this bit-sliced NTT design to construct a data-redundant countermeasure for NTT to provide fault-attack resistance. We have implemented our bit-sliced NTT/INTT design and integrated it with reference Dilithium implementation in C. In this section, we describe the bit-sliced code generation process, the bit-sliced design of key components of Dilithium and the bit-sliced dual data-redundant countermeasure design for NTT/INTT.

\subsection{Bit-sliced code generation process} \label{codegen}
Bit-slicing requires a boolean program (bit-level program), which is difficult to program because only single-bit operations can be used. To simplify the development of this boolean program, we express its equivalent functionality at the Register-Transfer Level (RTL) in a hardware description language, and then use logic synthesis to convert this behavior into an equivalent description of boolean operations. We have automated this step into an open-source logic synthesis environment that creates a bit-slice program in C directly from an RTL description in Verilog \cite{psp}.

\begin{figure}[t!]
    \centering
    \includegraphics[width=0.5\linewidth]{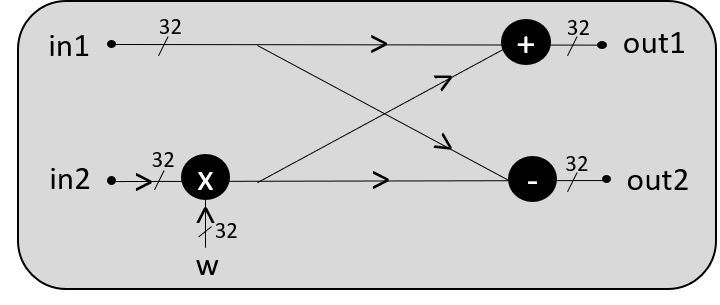}
    \caption{\centering{Cooley-Tukey Butterfly Unit in Verilog}}
    \label{fig:butterfly}
    \vspace{-6mm}
\end{figure}

Figure \ref{fig:butterfly} shows register-transfer level description of standard Cooley-Tukey butterfly unit. This Butterfly unit forms the foundation block of NTT transforms and is built from a $32$-bit modular adder, a $32$-bit modular subtractor and a $32$-bit modular multiplier module with modular operations in $\mathbb{F}_q$, where, prime modulus parameter $q$ is fixed to $2^{23} - 2^{13} + 1$ ($23$ bits) for Dilithium algorithm. Modular reduction in adder and subtractor is performed using conditional subtraction and addition respectively. While design of modular reduction in multiplier is based on barret reduction algorithm \cite{crypto-1986-1000} as described in Algorithm 6 \cite{Banerjee_Ukyab_Chandrakasan_2019}. Butterfly operation takes three $32$-bit signed integers as input: $in1$, $in2$ and twiddle factor $w$. The Butterfly operation then computes $in1 + (in2*w) \, \text{mod} \, q$ and $in1 - (in2*w) \, \text{mod} \, q$ resulting in two outputs $out1$ and $out2$ respectively. 

The Verilog RTL description for the butterfly unit is first converted to a boolean program, a gate-level netlist comprised of primitives such as \texttt{AND, OR, NOT} and \texttt{XOR} using logic synthesis. This boolean program is further converted into a C code by levelling the gates in the netlist and each gate is substituted with an equivalent bit-wise logical instruction. This C code generation is integrated as a backend pass in \texttt{YOSYS}, the open-source synthesis tool \cite{Yosys}.
Listing~\ref{lst:bit_sliced_code} illustrates the generated bit-sliced C code.  Each input or output in Figure \ref{fig:butterfly} is mapped into a
corresponding \texttt{int32\_t} representing 32 parallel slices. The input argument $state$ is passed a zero value since the butterfly unit is essentially a combinational circuit thus requires no state preservation across multiple calls to bit-sliced function.
To execute this bit-sliced code, the inputs must be converted into bit-sliced format. This requires doing matrix transposition of 32 $n$-bit inputs into $n$ 32-bit registers, for example, $in1[28]$ holds the $28^{th}$ bit of 32 different inputs. Once the bit-sliced computation is done, outputs are reverse-transposed from bit-sliced format into normal format.  

\begin{lstlisting}[float=*, language=C, frame=tb, tabsize=2, basicstyle=\small, caption=Generated Bit-sliced C code for Butterfly Unit, label={lst:bit_sliced_code}]
 1: void butterflyCompute(int32_t in1[32], int32_t in2[32], int32_t w[32]
    , int32_t out1[32], int32_t out2[32], int32_t state) {
 2:	    int32_t n00620_;
 3:     int32_t n00631_;
 4:     int32_t n00642_;
 5:     int32_t n00653_;
 6:     int32_t n08793_;
 7:     int32_t n07528_;
 8:     int32_t n07561_;
 9:     int32_t n08804_;
10:     int32_t q[23];
11:     ...
12:     NOT1(w[0], n00620_);
13:     NOT1(in2[0], n00631_);
14:     NOT1(w[18], n00642_);
15:     NOT1(in1[28], n00653_);
16:     AND2(in2[0], w[21], n08793_);
17:     OR2(n07528_, n07561_, n08804_);
18:     ...
19:     }
\end{lstlisting}

\subsection{Bit-sliced Components of Dilithium}
\label{section:bit-sliced_components}
In this section, we describe the mapping of key building blocks of Dilithium such as forward and inverse NTT transformations (NTT and INTT), polynomial multiplication and matrix-vector Polynomial Multiplication to bit-sliced code. We target bit-slicing to those sections of code of the Dilithium algorithm which are both computationally intensive and vulnerable with respect to fault attacks \cite{bruinderink2018differential, ravi2018side, ravi2019exploiting}, side-channel attacks \cite{10.1007/978-3-319-66787-4_25, 10.1007/978-3-030-30530-7_7} and cache-timing attacks \cite{8494855}. Bit-slicing provides constant-time implementation, enables intra-instruction redundancy to protect against fault-attacks and higher-order masking to protect against power side-channels \cite{10.1007/978-3-030-68773-1_11}. Please refer Algorithms ~\ref{alg:protected_dilithium_sign} and  \ref{alg:protected_dilithium_key} where the FA protected polynomial multiplication and matrix-vector polynomial multiplication instances are highlighted in blue.

\begin{algorithm}
    \caption{Polynomial multiplications within the Dilithium Signature Generation \cite{ducas2018crystals} algorithm which are protected using the bit-slicing based FA countermeasures are highlighted in blue.}
    \label{alg:protected_dilithium_sign}
    \begin{algorithmic}[1]
    \State \textbf{Input:} $sk$ - Secret Key, $M$ - Message
    \State \textbf{Output:} $\sigma$ - Signature
    \State $\textbf{A} \in R_{q}^{k \times l} \gets ExpandA(\rho)$
    \State $\mu \in \{0, 1\}^{384} \gets CRH(tr \parallel M)$
    \State $\kappa \gets 0, (z, h) \gets \perp$
    \State $\rho' \in \{0, 1\}^{384} \gets CRH(K \parallel \mu)$ (or $\rho' \gets \{0, 1\}^{384}$ randomized)
    \While{$(z, h) = \perp$}
        \State $y \in S_{\gamma1}^l \gets ExpandMask(\rho', \kappa)$
        \State $w \gets \textcolor{blue} {\textbf{A}y}$ \Comment{\textcolor{blue}{$ {\textbf{A}y}$ is protected}}
        \State $w_{1} \gets HighBits_{q}(w, 2\gamma_{2})$
        \State $\tilde{c} \in \{0, 1\}^{256} \gets H(\mu \parallel w_{1})$
        \State $c \in B_{\tau} \gets SampleInBall(\tilde{c})$
        \State $z \gets y+\textcolor{blue}{c.s_{1}}$  \Comment{\textcolor{blue}{$c.s_{1}$ is protected}} \label{lst:line:z}
        \State $r_{0} \gets LowBits_{q}(w - \textcolor{blue}{c.s_{2}}, 2\gamma_{2})$ \Comment{\textcolor{blue}{$c.s_{2}$ is protected}}
        \If {$\Vert z\Vert \geq \gamma_{1} - \beta$ or $\Vert r_{0}\Vert_{\infty} \geq \gamma_{2} - \beta$} 
            \State $(z, h) \gets \perp$ \label{lst:line:rejection_sampling}
        \Else
            \State $h \gets MakeHint_{q}(-\textcolor{blue}{c.t_{0}}, w - \textcolor{blue}{c.s_{2}} + \textcolor{blue}{c.t_{0}}, 2\gamma_{2})$ \Comment{\textcolor{blue}{$c.t_{0}$ and $c.s_{2}$ are protected}}
            \If {$\Vert \textcolor{blue}{c.t_{0}}\Vert_{\infty} \geq \gamma_{2}$ or the \# of 1's in $h>\omega$} \Comment{\textcolor{blue}{$c.t_{0}$ is protected}}
                \State $(z, h) \gets \perp$
            \EndIf
        \EndIf
    \State $\kappa \gets \kappa + l$
    \EndWhile
    \State \textbf{return} $\sigma = (z, h, \tilde{c})$
    \end{algorithmic}
\end{algorithm}

\begin{algorithm}
    \caption{Polynomial multiplication within the Dilithium Key Generation \cite{ducas2018crystals} algorithm which are protected using the bit-slicing based FA countermeasures are highlighted in blue.}
    \label{alg:protected_dilithium_key}
    \begin{algorithmic}[1]
    \State \textbf{Output:} $pk$ - Public Key, $sk$ - Secret Key
    \State $\zeta \gets \{0,1\}^{256}$
    \State $(\rho, \varsigma, K) \in \{0,1\}^{256 \times 3} \gets H(\zeta)$
    \State $(s_1, s_2) \in {S_\eta^l \times S^k_\eta} \gets H(\varsigma)$
    \State $A \in R_q^{k \times l} \gets ExpandA(\rho)$
    \State $t \gets \textcolor{blue}{A s_1} + s_2$ \Comment{\textcolor{blue}{$As_{1}$ is protected}}
    \State $(t_1, t_0) \gets Power2Round_q(t, d)$
    \State $\textit{tr} \in \{0,1\}^{384} \gets CRH(\rho \Vert t_1)$
    \State \textbf{return} $(pk = (\rho, t_1), sk = (\rho, K, \textit{tr}, s_1, s_2, t_0))$

 \end{algorithmic}
\end{algorithm}

\subsubsection{Bit-sliced NTT and Inverse NTT}\hfill

We exploit the iterative nature of NTT algorithm to apply fine-grained parallelism through bit-slicing. The NTT/INTT of size $n$ is typically computed in total $\log(n)$ stages with each stage consisting of $n/2$ butterfly operations. Dilithium algorithm operates with polynomials of size $n = 256$ and uses 256-point NTT and 256-point INTT to compute polynomial multiplication. Cooley-Tukey (CT) butterfly structure is used to perform both NTT and INTT operations. The CT butterfly-based NTT and INTT requires inputs in bit-reversed order and generates outputs in normal order, similar to the decimation-in-time (DIT) FFT. While the INTT operates similar to NTT expect that its twiddle factors are negative powers of $\omega$  and coefficient-wise multiplication of output by $n^{-1}$ and powers of $\psi^{-1}$ as defined in Equations \ref{eq:NTT_formula} and \ref{eq:INTT_formula}. In order to map 256-point NTT to bit-sliced mode, 256-point NTT is broken down into four 64-point NTTs. Now, each 64-point NTT computation takes six stages and each stage has 32 butterfly operations. Since all butterflies within any stage of the NTT can be computed independent of one another, this allows us to apply bit-sliced approach to NTT architecture. To parallelize the 32 butterfly operations in any stage of 64-point NTT, it is computed using a single call to bit-sliced butterfly function \textsc{butterflyCompute()} as defined in Listing~\ref{lst:bit_sliced_code}.

\begin{figure*}[t!]
    \centering
    \includegraphics[width=1.0\columnwidth]{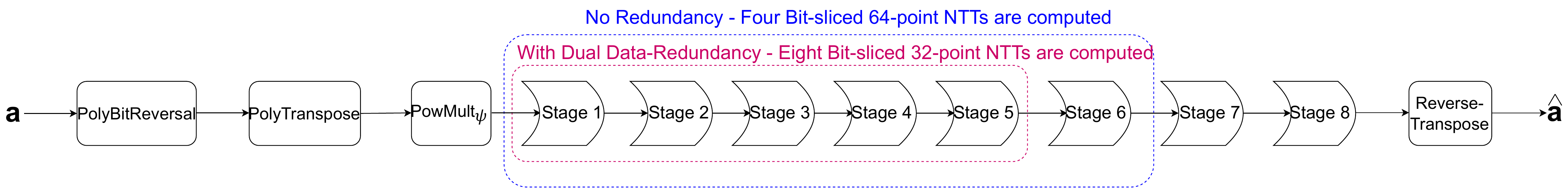}
    \caption{\raggedright{Block diagram for Bit-sliced 256-point NTT.}}
    \label{fig:bitsliced256pointNTT}
\end{figure*}

\begin{figure*}[t!]
    \centering
    \begin{subfigure}[t]{0.45\textwidth}
        \centering
        \includegraphics[width=\textwidth]{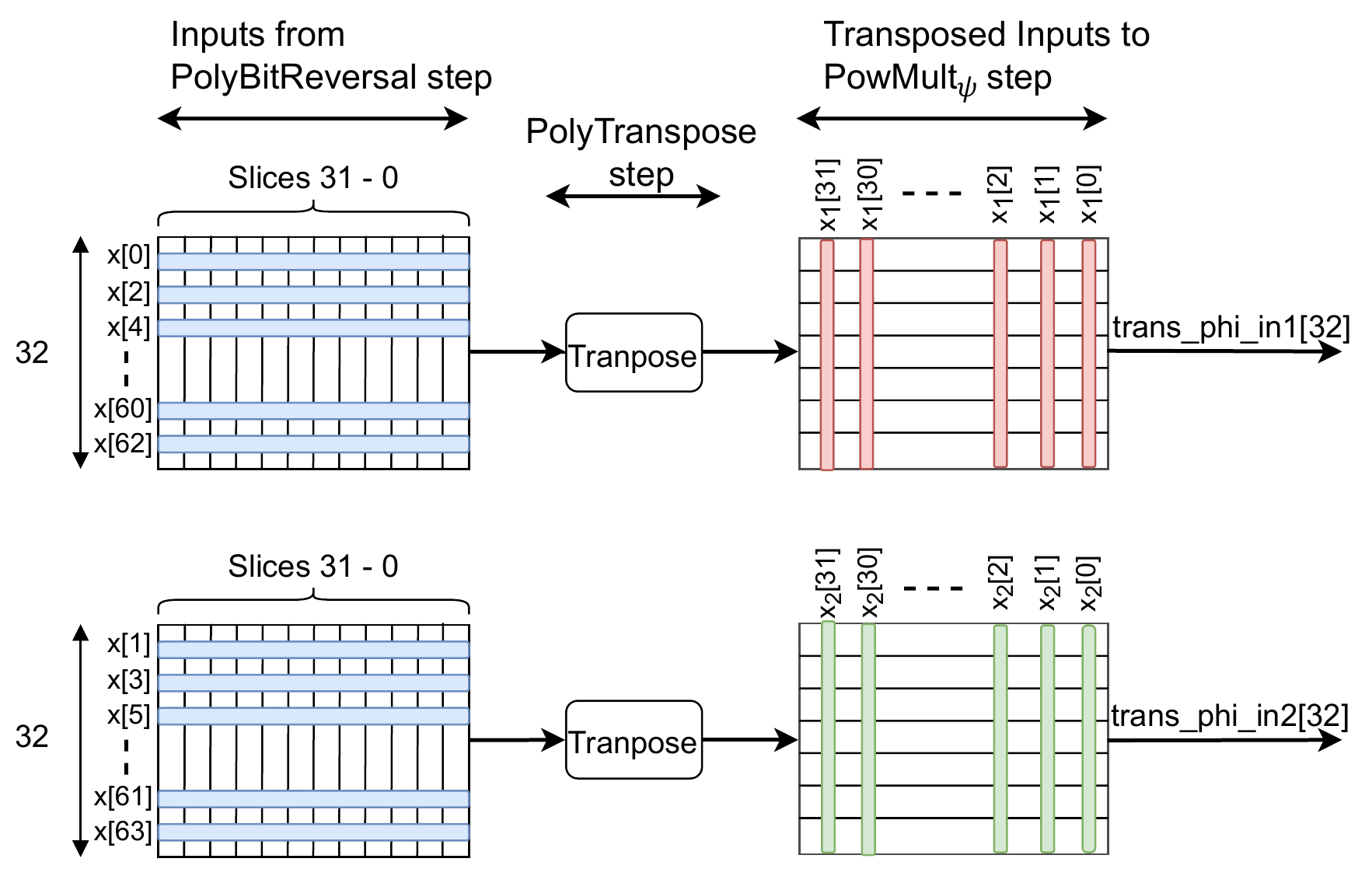}
        \caption{Bit-slicing format}
        \label{fig:bitslicing_format}
    \end{subfigure} %
    ~ 
    \begin{subfigure}[t]{0.45\textwidth}
        \centering
        \includegraphics[width=\textwidth]{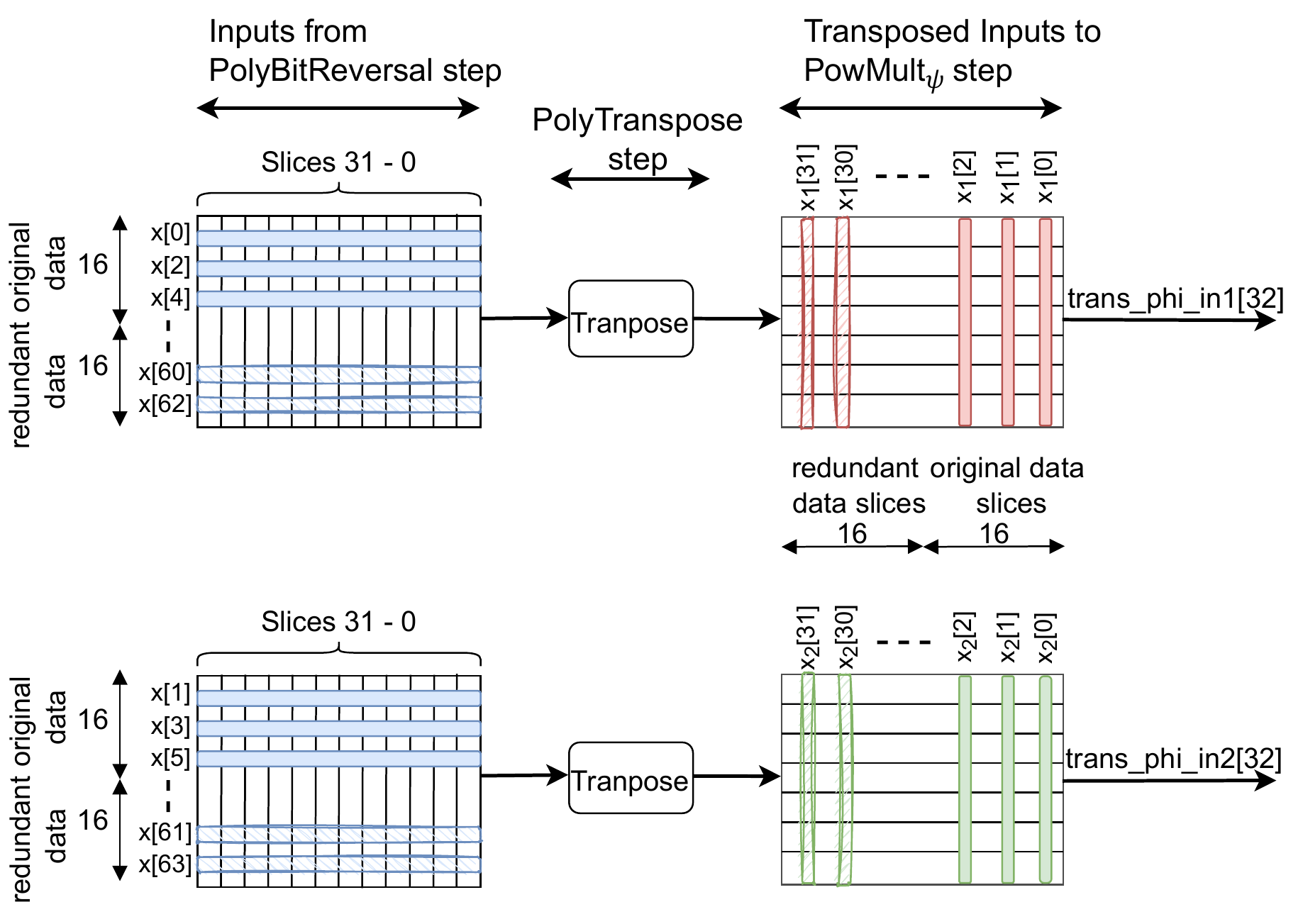}
        \caption{Dual data-redundancy bit-slicing format}
        \label{fig:bitslicing_dualdataredundancy_format}
    \end{subfigure} %
    ~
        \caption{Representation of NTT polynomial coefficients in bit-slicing format without and with dual data-redundancy.}
\end{figure*}

Figure \ref{fig:bitsliced256pointNTT} shows the step-wise transformation of an input polynomial $\textbf{a}$ into 256-point NTT output polynomial $\hat{\textbf{a}}$ using bit-slicing technique, whereas, Algorithm ~\ref{alg:256_NTT_bitsliced} outlines its pseudocode. In lines~\ref{lst:line:start_bitsliced_manipulation}-\ref{lst:line:end_bitsliced_manipulation} of the pseudocode, \textsc{bit-sliced manipulation} process takes place using SIMD instructions explained later in the section, we use ARM NEON intrinsics to support for Cortex-A series processors. This particular section of code can easily be ported to support AVX2 SIMD instructions for Intel CPUs as well. The NTT transformation starts with \textsc{PolyBitReversal} step which  performs bit reversal on the input polynomial $a$:

\begin{eqnarray*}
b[i] &=& \text{PolyBitReversal}(a)[i] = a[\text{BitReversal}(i)] \\
\text{where, BitReversal}(i) &=& \sum_{j=0}^{nS-1} {(((i \gg j) \And 1) \ll (nS - 1 - j))} \\
\text{and nS} &=& \log(n)
\end{eqnarray*}

\begin{figure*}[t!]
    \centering
    \includegraphics[width=1.0\columnwidth]{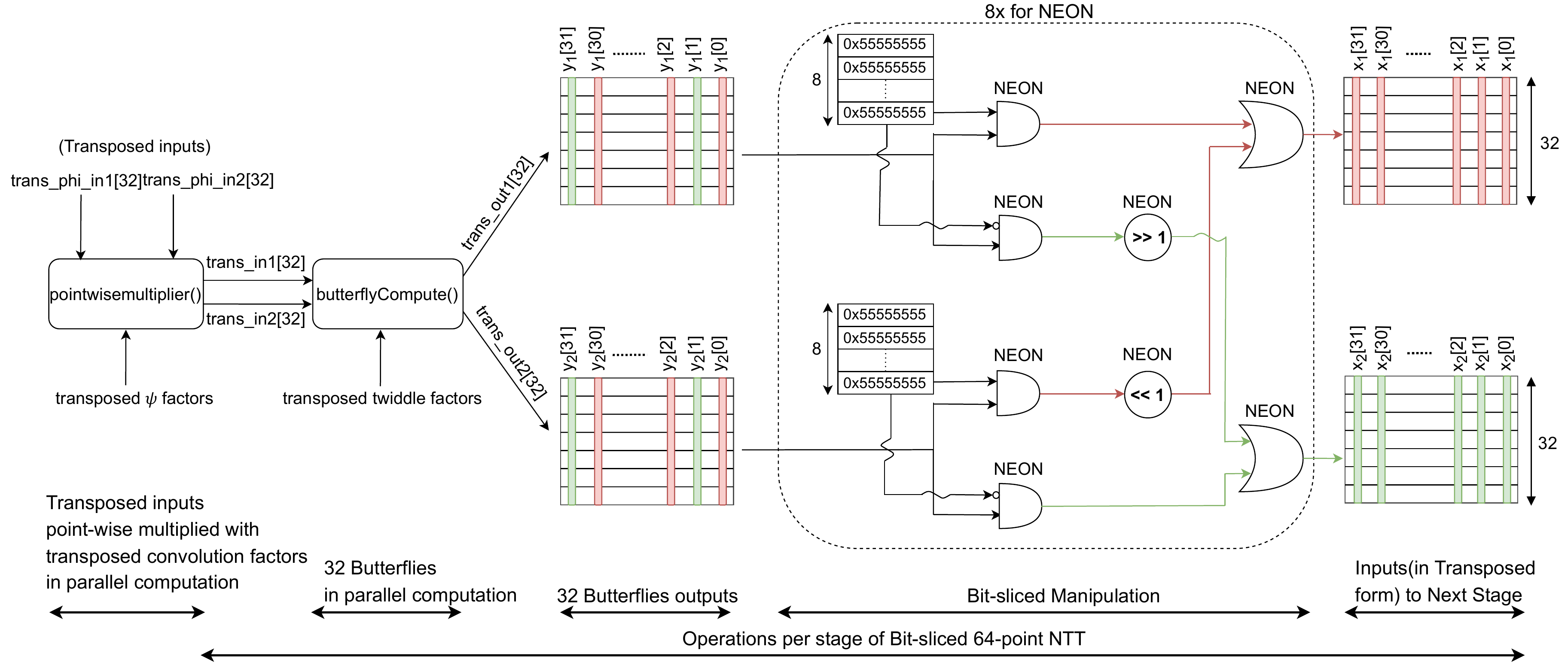}
    \caption{\raggedright{Data flow diagram of Bit-sliced 256-point NTT composed of four 64-point NTTs. The figure shows operations that occur in each stage of a Bit-sliced 64-point NTT.}}
    \label{fig:bitsliced256pointNTT_Part2}
\end{figure*}

\begin{figure*}[t!]
    \centering
    \includegraphics[width=1.0\columnwidth]{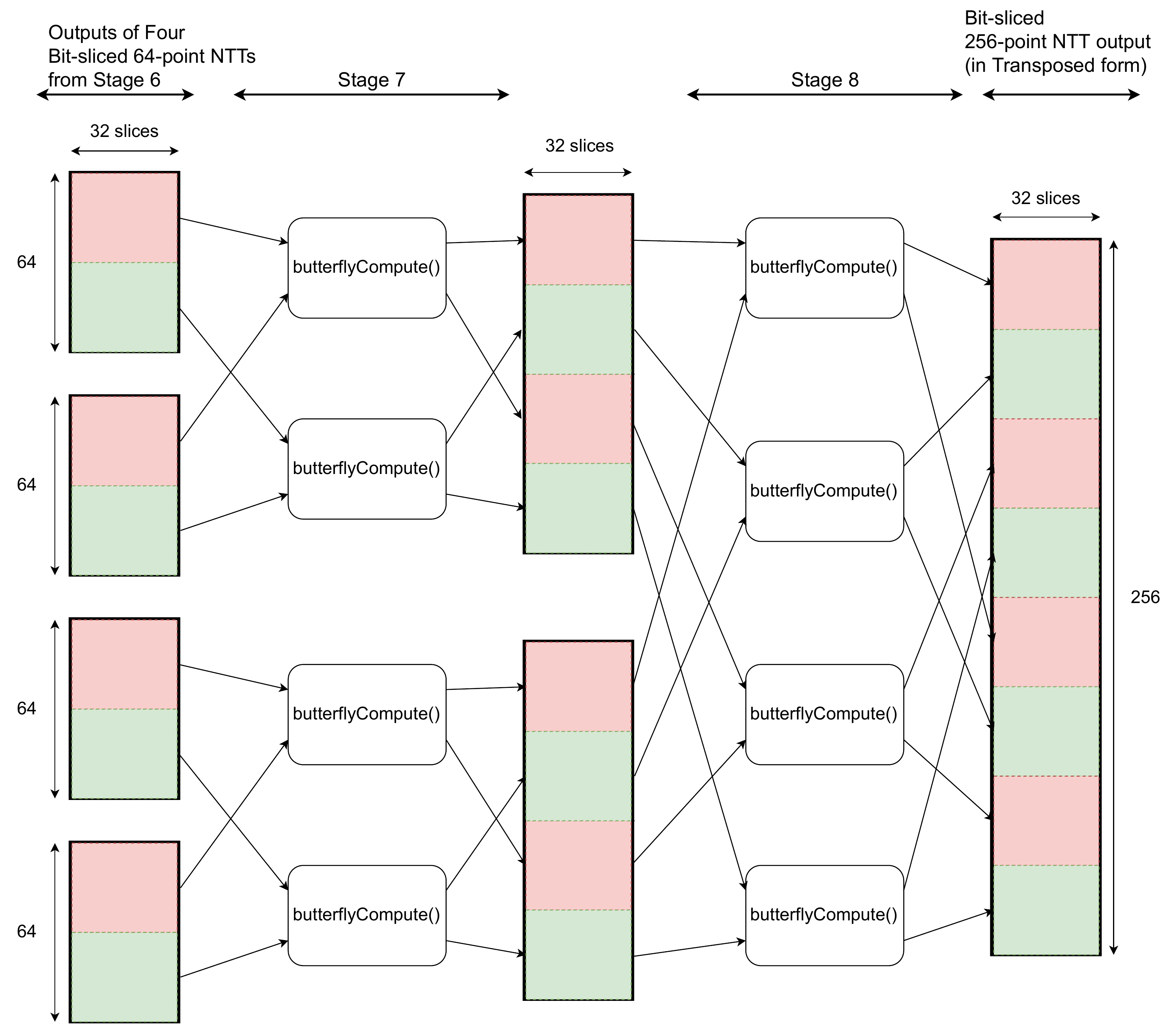}
    \caption{\raggedright{Data flow diagram in Stages 7 and 8 of Bit-sliced 256-point NTT. A bit-sliced 64-point NTT output from the $6^{th}$ stage of 256-point NTT is represented using a rectangular block with top-half/red portion occupied with 32 first outputs (in 32 adjacent red slices spread across 32 words) and bottom-half/green portion occupied with 32 second outputs (in 32 adjacent green slices spread across 32 words) of 32 butterflies.}}
    \label{fig:bitsliced256pointNTT_Part1}
\end{figure*}

The bit-reversed input polynomial is then bisected and stored into arrays $in1$ and $in2$ as the two set of inputs applied to the 128 butterflies in the first stage of 256-point NTT as shown in lines~\ref{lst:line:start_store_poly} - \ref{lst:line:end_store_poly} of code. Each array represents 4 subsets each of size $n/8$ and one subset drawn from each array therefore forms the 64 inputs to a 64-point NTT block. In other words, first subset and second subset constitutes the first inputs and second inputs respectively for the 32 butterflies of a 64-point NTT. These two butterfly inputs subsets can be viewed as $x[0], x[2], x[4], \dotsc, x[62]$ and $x[1], x[3], x[5], \dotsc, x[63]$ respectively in the Figure \ref{fig:bitslicing_format}. Before $32$-bit input polynomial coefficients can be processed by a bit-sliced function, the input has to be transposed. The \textsc{PolyTranspose} step in Figure \ref{fig:bitsliced256pointNTT} performs transpose on the two butterfly inputs arrays and converts $n/8$ 32-bit inputs per first and second subset into 32 $n/8$-bit inputs each as depicted by red and green colored slices respectively in the Figure \ref{fig:bitslicing_format}. In Figure \ref{fig:bitsliced256pointNTT_Part2}, the transposed butterfly inputs subsets $trans\_psi\_in1$ and $trans\_psi\_in2$ are then passed as input arguments to \textsc{pointwisemultiplier()} bit-sliced function at \textsc{PowMul}${_\psi}$ step of Figure \ref{fig:bitsliced256pointNTT} to realize multiplication of polynomial coefficients with respective powers of negative convolution ($\psi$) factors. Its output subsets $trans\_in1$ and $trans\_in2$  are then inputs to \textsc{butterflyCompute()} bit-sliced function resulting in two butterfly outputs subsets, $trans\_out1$ and  $trans\_out2$, at the same time, comprising of the 32 first outputs and 32 second outputs respectively for the 32 butterflies in $1^{st}$ stage of a 64-point NTT block. In Figure \ref{fig:bitsliced256pointNTT_Part2}, the 32 alternate red and green slices in two butterfly outputs subsets represents intermediate outputs which become the first inputs and second inputs respectively to the $2^{nd}$ stage butterflies of a 64-point NTT. In order to process $2^{nd}$ stage inputs by \textsc{butterflyCompute()} function, all the first butterfly inputs represented by red slices should be placed together in one subset and all the second butterfly inputs represented by green slices should to be placed together in the other subset. We therefore perform \textsc{bit-sliced manipulation} operation after bit-sliced butterfly computation operation in first five stages of 64-point NTT to obtain the desired butterfly inputs subsets for next stage as shown in the Figure \ref{fig:bitsliced256pointNTT_Part2}. We exploit the fact that NTT is easily vectorizable hence \textsc{bit-sliced manipulation} takes the two input subsets through a sequence of \texttt{SIMD} bit-wise operations as shown in the lines~\ref{lst:line:start_bitsliced_manipulation}-\ref{lst:line:end_bitsliced_manipulation} of pseudocode and can also be perceived as a gate-level circuit illustrated in Figure \ref{fig:bitsliced256pointNTT_Part2}. This process begins with masking the inter-mixed red slices and green slices in both the subsets with a given mask value and its complement respectively. To extract and set together the red slices from both the subsets, the \texttt{SIMD} ANDed output of red slices from second subset with a complementary mask value is shifted by a given shift value and finally, it is \texttt{SIMD} ORed with the \texttt{SIMD} ANDed output of red slices from first subset with mask value. The results of \textsc{bit-sliced manipulation} are stored back into same memory locations from where butterfly inputs are read. Mask value, $bitshufflemask$, complementary mask value, $bitshuffleinmask$ and shift value, $shiftval$ are configured for first five stages of 64-point NTT stage since the number of overlapping butterflies differs stage-wise. Figure \ref{fig:bitsliced256pointNTT_Part2} highlights the path with red and green which combines all red slices together and green slices together respectively. Last stage of 64-point NTT does not require \textsc{bit-sliced manipulation} operation because all the 32 butterflies are interleaved with each other which causes both the butterfly output subsets to already have all the first butterfly outputs in one subset and all the second butterfly outputs in the other subset. A 64-point NTT output in bit-sliced domain is represented in Figure \ref{fig:bitsliced256pointNTT_Part1} using a rectangular block with top-half/red portion occupied with 32 first outputs (in 32 adjacent red slices spread across 32 words) and bottom-half/green portion occupied with 32 second outputs (in 32 adjacent green slices spread across 32 words) of 32 butterflies in the $6^{th}$ stage.

To compute the $7^{th}$ stage of 256-point NTT, 4 64-point NTT blocks are combined to form 2 128-point NTT blocks by 4 invocations to \textsc{butterflyCompute()} bit-sliced function as in lines~\ref{lst:line:start_seventh_stage}-\ref{lst:line:end_seventh_stage} of pseudocode. Figure \ref{fig:bitsliced256pointNTT_Part1} shows that red portions and green portions of top two 64-point NTT blocks are inputs for the first 32 butterflies and next 32 butterflies respectively in the 7th stage of 256-point NTT which combine together to produce first 128-point NTT block. Butterfly outputs are stored back into same memory locations where the inputs are in. Going to the $8^{th}$ stage, composition of 2 128-point NTT blocks leads to a single 256-point NTT block which is computed by 4 invocations to \textsc{butterflyCompute()} bit-sliced function as in lines~\ref{lst:line:start_eighth_stage}-\ref{lst:line:end_eighth_stage} of pseudocode. We get the final 256-point NTT output by performing last step in Figure \ref{fig:bitsliced256pointNTT}, \textsc{ReverseTranspose} to convert the 32-bit 256-point NTT outputs from bit-sliced domain to normal domain. 

It is to be noted that we use pre-computed transposed twiddle factors: $\omega$, ${\omega}^{-1} \text{mod} \, q$ and $\psi$, $n^{-1}{{\psi}^{-1}} \text{mod} \, q$ in our bit-sliced implementation of 256-point NTT and INTT. The bit-sliced implementation for 256-point INTT is same as 256-point NTT except that there is \textsc{PowMul}${_\psi}$ step at the end before the \textsc{ReverseTranspose} step instead of the beginning to perform point-wise multiplication of INTT output polynomial coefficients with respective powers of $n^{-1}{{\psi}^{-1}}$ using \textsc{pointwisemultiplier()} bit-sliced function.



\subsubsection{Bit-sliced Polynomial Multiplication}
An NTT-based multiplication of two polynomials is computed using two forward NTTs to transform the input polynomials, a point-wise vector multiplication and one inverse NTT as stated in Equation ~\ref{eq:Polymult_formula}. Bit-sliced implementation of multiplication of two 256 elements polynomials uses bit-sliced instances of: 256-point NTT and 256-point INTT for forward and inverse transform; \textsc{pointwisemultiplier()} bitsliced function to perform point-wise multiplication of 2 polynomials in NTT domain representation. 

\subsubsection{Bit-sliced Matrix-Vector Polynomial Multiplication}
Bit-sliced multiplication of a $M \times N$ polynomials matrix   with a $N \times 1$ polynomials vector is computed by calculating the inner product row-wise and the inner product is composed of $N$ bit-sliced polynomial multiplications and $N$ bit-sliced polynomial accumulations. \textsc{pointwiseaccumulator()} bit-sliced function uses a state element to accumulate over the $N$ polynomials resulting from the point-wise multiplication of $N$ NTT domain polynomials in a given row of a matrix with $N$ NTT domain polynomials of a vector. This state element is mapped into a global variable to preserve state across multiple invocations of \textsc{pointwiseaccumulator()} function. To accumulate $N$ polynomials, we run bit-sliced accumulation over their 32 coefficients using $N$ calls to \textsc{pointwiseaccumulator()} and by iterating bit-sliced accumulation over the size of the polynomial, it outputs a single polynomial in bit-sliced format which is then followed by a bit-sliced INTT transformation to get the final polynomial for that particular row.

\subsection{Proposed Fault-Attack Countermeasure for Dilithium Algorithm}
\label{section:proposed_countermeasure}
In this section, we describe the implementation of Fault-Attack Countermeasure based on IIR mechanism in Dilithium. We utilize the bit-sliced construction defined for key elements of Dilithium in Section~\ref{section:bit-sliced_components} as the foundation to add dual spatial redundancy to it. We consider the case of 256-point NTT to demonstrate the IIR-based countermeasure in Dilithium. In bit-sliced implementation of 256-point NTT, each slice is allocated to operate on a different coefficient of a polynomial for maximum throughput. We separate slices into original data slices (ODS) and redundant data slices (RDS) occupying the lower-half and upper-half word respectively for fault detection. RDS and ODS operate on the same input coefficients at any intermediate stage of 256-point NTT, and thus, they produce same output if no fault occurs. If a fault occurs during their execution, then it will be detected when output polynomial coefficients are compared at the end of 256-point NTT transformation.

\begin{figure*}[t!]
    \centering
    \includegraphics[width=1.0\columnwidth]{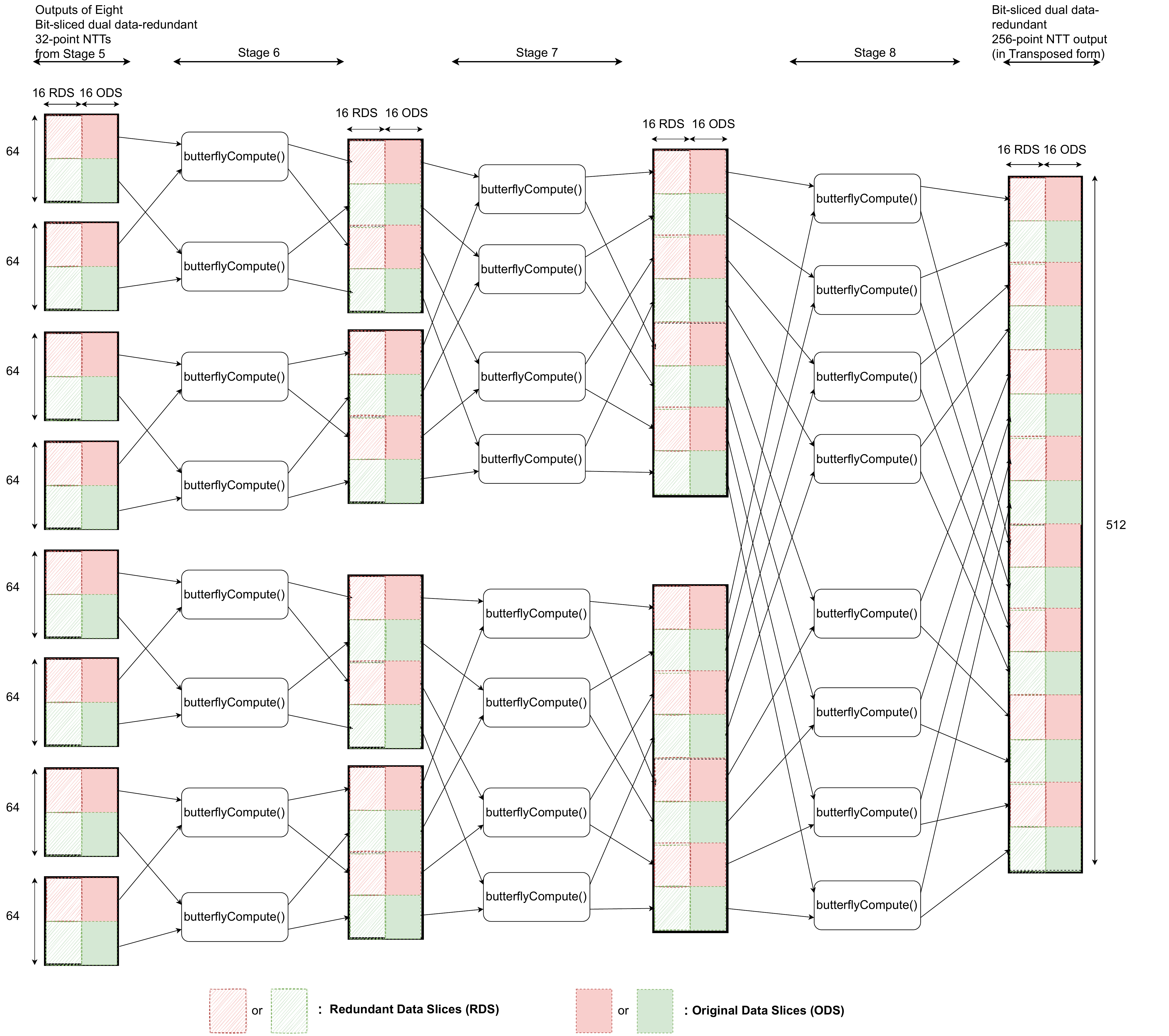}
    \caption{\raggedright{Data flow diagram in stages 6, 7 and 8 of Bit-sliced 256-point NTT with dual data-redundancy}}
    \label{fig:bitslicedwithRedundancy256pointNTT}
\end{figure*}

Polynomial coefficients array of size $n$ arranged in bit-reversed order is split into 16 subsets each of size $n/8$ such that each subset has first half elements as polynomial coefficients and remaining are its copies. While half of the subsets store coefficients which are first inputs and remaining half subsets have second inputs of $n/2$ butterflies in first stage of 256-point NTT. Then, at \textsc{PolyTranspose} step, subsets are transposed into bit-sliced format having dual-spatial redundancy. Figure~\ref{fig:bitslicing_dualdataredundancy_format} illustrates a part of \textsc{PolyTranspose} step, when one of the first butterfly inputs subsets and one from the second butterfly inputs subsets, both shown in blue, are converted into their transposed versions represented by red and green respectively wherein the redundant slices and original data slices are portrayed in different patterns and indicated by arrows. All slices are split across 32 words for 32-bit polynomial coefficients. In next step, transposed subsets are point-wise multiplied with their respective power of $\psi$ factors using \textsc{pointwisemultiplier()} bit-sliced function. 

Next, when a first butterfly inputs subset and a second butterfly inputs subset are passed to \textsc{butterflyCompute()} bit-sliced function, half of the butterflies will operate on redundant inputs, thus produces two butterfly output subsets with redundant slices which then become inputs to next stage. This leads us to compute in-parallel two 32-point NTTs, original and its copy, with 16 butterflies and its copies being computed per stage using a single call to \textsc{butterflyCompute()}. Here also, each stage except the last one involves \textsc{bit-sliced manipulation} process after \textsc{butterflyCompute} similar to non-redundant bit-sliced implementation. With dual data-redundancy, bit-sliced 256-point NTT is broken down into 8 32-point NTT blocks instead of 4 64-point NTT blocks. These 8 32-point NTT blocks are computed on 8 different pairs of butterfly inputs subsets and each 32-point NTT output in transposed form is represented in Figure~\ref{fig:bitslicedwithRedundancy256pointNTT} using a rectangular block with top-half/red portion and bottom-half/green portion storing first butterfly outputs and second butterfly outputs respectively of last stage of a 32-point NTT block. So, $6^{th}$ stage of 256-point NTT will merge these 8 32-point NTT blocks into 4 64-point NTT blocks using 8 calls to \textsc{butterflyCompute()} function in a way as illustrated in Figure~\ref{fig:bitslicedwithRedundancy256pointNTT}. In $7^{th}$ stage, 4 64-point NTTs combine to output 2 128-point NTT blocks. 2 128-point blocks combine to output final 256-point NTT output. It is to be noted that in each stage, NTT blocks are generated along with its copies in-parallel. 

An adversary can bypass this countermeasure by injecting two single-bit faults that are in different halves of the processor word. These two faults have to align with any of the ODS and its respective RDS. Then, both will produce same faulty 256-point NTT output, going undetected.

\begin{algorithm*}
   
    \caption{Bit-sliced 256-point NTT}
    \label{alg:256_NTT_bitsliced}
    \begin{algorithmic}[1]
            \State \textbf{Input}: $a(x) \in R_q$ 
            \State \textbf{Output}: $\hat{a}(x) \in R_q$ \text{such that} $\hat{a} = \text{NTT}(a)$   
            \State \textbf{Setup}: Pre-computed transposed twiddle factors $trans\_w$ and transposed $\psi$ factors $trans\_psi1$ and $trans\_psi2$
            \State $state1 \gets 0, state2 \gets 0$     \Comment{Initialization} \\
            $bitshufflemask[5] \gets \{\texttt{0x55555555}, \, \texttt{0x33333333}, \, \texttt{0x0F0F0F0F}, \, \texttt{0x00FF00FF}, \, \texttt{0x0000FFFF}\} $  \\
            $bitshuffleinmask[5] \gets \{\texttt{0xAAAAAAAA}, \texttt{0xCCCCCCCC}, \texttt{0xF0F0F0F0}, \texttt{0xFF00FF00}, \texttt{0xFFFF0000}\} $
            \State $b \gets \text{PolyBitReversal}(a)$  \Comment{Polynomial Bit-Reversal}
            \For{$i  \gets 0$ \textbf{to} $n/2$}     \label{lst:line:start_store_poly}\Comment{Store polynomial in butterfly inputs form}
                \State $in1[i] \gets b[i*2]$          
                \State $in2[i] \gets b[i*2 + 1]$
            \EndFor \label{lst:line:end_store_poly}
            \For{$i  \gets 0$ \textbf{to} $n/64$}     \Comment{Transpose butterfly inputs polynomial to bit-sliced format}
                \State $trans\_psi\_in1[i*32] \gets \text{Transpose}(in1[i *32])$
                \State $trans\_psi\_in2[i*32] \gets \text{Transpose}(in2[i *32])$
            \EndFor
            
            \For{$i  \gets 0$ \textbf{to} $n/64$}  
            \Comment{Point-wise multiply butterfly inputs polynomial with respective negative convolution factors}
                \State $\textsc{pointwisemultiplier}(\&trans\_psi\_in1[i*32], \, \&trans\_psi1[i*32], \, \&trans\_in1[i*32], \newline \, state2)$
                \State $\textsc{pointwisemultiplier}(\&trans\_psi\_in2[i*32], \, \&trans\_psi2[i*32], \, \&trans\_in2[i*32], \newline \, state2)$
            \EndFor
            
            \For{$j  \gets 0$ \textbf{to} $n/64$}    \Comment{Loop over 4 64-point NTTs}
                \State $num \gets 0$ 
                \While{$num < 6$}                    \Comment{Loop over 6 stages of 64-point NTT}
                    \State $\textsc{butterflyCompute}(\& trans\_in1[j*32], \, \& trans\_in2[j*32],$  \\\hspace{0.9cm} 
                    $\& trans\_w[j*32], \, \& trans\_out1[j*32], \, \& trans\_out2[j*32], \, state1)$  \Comment{Bit-sliced computation for 32 butterflies} 
                    
                    \If{$num \, != 5$}  
                    \label{lst:line:start_bitsliced_manipulation}
                    \Comment{Bitsliced manipulation to configure butterfly inputs for next stage}
                        \State $mask \gets \text{vdupq\_n\_u32}(bitshufflemask[num])$
                        \State $invmask \gets \text{vdupq\_n\_u32}(bitshuffleinmask[num])$
                        \State $shiftval \gets 2^{num}$
                        \For{$i \gets 0$ \textbf{to} $8$} \Comment{Iterate over 32 butterfly outputs in bit-sliced format}
                            \State $out1 \gets \text{vld1q\_u32}(\& trans\_out1[i*4 + j*32])$
                            \State $out2 \gets \text{vld1q\_u32}(\& trans\_out2[i*4 + j*32])$
                            \State $andout1 \gets \text{vandq\_u32}(out1, mask)$
                            \State $andout2 \gets \text{vandq\_u32}(out2, mask)$
                            \State $andnout1 \gets \text{vandq\_u32}(out1, invmask)$
                            \State $andnout2 \gets \text{vandq\_u32}(out2, invmask)$
                            \State $lsout \gets \text{vshlq\_n\_u32}(andout2, shiftval)$
                            \State $orout1 \gets \text{vorrq\_u32}(andout1, lsout)$
                            \State $rsout \gets \text{vshrq\_n\_u32}(andnout1, shiftval)$
                            \State $orout2 \gets \text{vorrq\_u32}(andnout2, rsout)$
                            \State $\text{vst1q\_u32}(\& trans\_in1[i*4 + j*32], orout1)$
                            \State $\text{vst1q\_u32}(\& trans\_in2[i*4 + j*32], orout2)$
\algstore{continue}            
\end{algorithmic}
\end{algorithm*}

\begin{algorithm*}
    \begin{algorithmic}[1]
    \algrestore{continue}
                        \EndFor
                    \EndIf
                    \label{lst:line:end_bitsliced_manipulation}
                \EndWhile
            \EndFor
            \For{$i  \gets 0$ \textbf{to} $n/2$ \textbf{by} $32$}        \Comment{Configure butterfly inputs for $7^{th}$ NTT stage}
                \For{$j  \gets 0$ \textbf{to} $32$}
                    \State $trans\_in1[j + i] \gets trans\_out1[j+i]$
                    \State $trans\_in2[j + i] \gets trans\_out2[j+i]$
                \EndFor
            \EndFor
            
             \Comment{Bit-sliced computation for 4 blocks with 32 butterflies/block at $7^{th}$ NTT stage}

            \State $\textsc{butterflyCompute}(\& trans\_in1[0], \, \& trans\_in1[32], \, \& trans\_w[num*32], \, \newline \& trans\_out1[0], \, \& trans\_out1[32], \, state1)$ \label{lst:line:start_seventh_stage}
            \State $\textsc{butterflyCompute}(\& trans\_in2[0], \, \& trans\_in2[32], \, \& trans\_w[(num+1)*32], \, \newline \& trans\_out2[0], \, \& trans\_out2[32], \, state1)$
            \State $\textsc{butterflyCompute}(\& trans\_in1[64], \, \& trans\_in1[96], \, \& trans\_w[num*32], \, \newline \& trans\_out1[64], \, \& trans\_out1[96], \, state1)$
            \State $\textsc{butterflyCompute}(\& trans\_in2[64], \, \& trans\_in2[96], \, \& trans\_w[(num+1)*32], \, \newline \& trans\_out2[64], \, \& trans\_out2[96], \, state1)$ \label{lst:line:end_seventh_stage}
            
            \For{$i  \gets 0$ \textbf{to} $n/2$ \textbf{by} $32$}   \Comment{Configure butterfly inputs for $8^{th}$ NTT stage}
                \For{$j  \gets 0$ \textbf{to} $32$}
                    \State $trans\_in1[j + i] \gets trans\_out1[j+i]$
                    \State $trans\_in2[j + i] \gets trans\_out2[j+i]$
                \EndFor
            \EndFor  
        \\\hspace{0.1cm} \Comment{Bit-sliced computation for 4 blocks with 32 butterflies/block at $8^{th}$ NTT stage}
            
        \State $\textsc{butterflyCompute}(\& trans\_in1[0], \, \& trans\_in1[64], \, \&  trans\_w[(num+1)*32], \, \newline \& trans\_out1[0], \, \& trans\_out1[64], \, state1)$ \label{lst:line:start_eighth_stage}
        
        \State $\textsc{butterflyCompute}(\& trans\_in2[0], \, \& trans\_in2[64], \, \& trans\_w[(num+2)*32], \, \newline \& trans\_out2[0], \, \& trans\_out2[64], \, state1)$
        \State $\textsc{butterflyCompute}(\& trans\_in1[32], \, \& trans\_in1[96], \, \& trans\_w[(num+3)*32], \, \newline \& trans\_out1[32], \, \& trans\_out1[96], \, state1)$
        \State $\textsc{butterflyCompute}(\& trans\_in2[32], \, \& trans\_in2[96], \, \& trans\_w[(num+4)*32], \, \newline \& trans\_out2[32], \, \& trans\_out2[96], \, state1)$ \label{lst:line:end_eighth_stage}

        \For{$i  \gets 0$ \textbf{to} $n/64$}     \Comment{Reverse Transpose NTT outputs polynomial}
            \State $out1[i*32] \gets \text{Transpose}(trans\_out1[i *32])$
            \State $out2[i*32] \gets \text{Transpose}(trans\_out2[i *32])$
        \EndFor
        \For{$i  \gets 0$ \textbf{to} $n/64$}   \Comment{Store outputs in final 256-point NTT output vector}
            \For{$j  \gets 0$ \textbf{to} $32$}
                \State $\hat{a}[j + 2*i*32] \gets out1[j+i*32]$
                \State $\hat{a}[j + (2*i + 1)*32] \gets out2[j+i*32]$
            \EndFor
        \EndFor
\end{algorithmic}
\end{algorithm*}        

\section{Fault Attacker Model}
\label{section:fault_attacker_model}
We assume that an attacker has complete physical access to the device during the execution of the signature generation procedure. The attacker should be able to collect the signatures generated by a message of his/her choice. The attacker should either be able to have a trigger in the device to time its fault injection or capability to capture power/EM traces to extract from them the timing information of points of interest. The attacker can stress the circuit or environmental conditions using an EM glitch. These glitches will induce localized transient faults into the device computation. An attacker can modify an instruction to be executed affecting the program flow of the algorithm or can cause algorithm modification depending upon the replaced instruction. An attacker can also modify the data to be loaded from memory without modifying the load instruction itself. 

Proy {\em et al.} investigated effects of EM-induced faults at Instruction Set Architecture (ISA) level on an ARM Cortex-A9 based SoC \cite{proysoc}. They have identified varied fault models ranging from instruction skip and register corruption, widely-adopted fault model for micro controllers \cite{7140238, ravi2019number}, to operand substitution, multiple correlated register corruptions and advanced control flow hijacking for SoCs. In this work, we have not performed characterization of fault effects rather assumed the studied fault effects on SoCs in previous works \cite{proysoc, thomassoc}.

\section{Security Analysis of the Proposed Countermeasure}
\label{section:security_analysis}
In this section, we provide the security analysis of the proposed countermeasure in Section~\ref{section:proposed_countermeasure} against the fault attacker models defined in Section~\ref{section:fault_attacker_model}.

In the unprotected, Dilithium implementation, any computation or instruction fault during the execution of signature generation algorithm may be useful for the adversary. As there is no detection mechanism for this implementation, the faulty signatures given out by the signature generation process may lead to some information about the secret key. The simplest method to protect against faults is function level temporal redundancy, i.e., to run the signature generation multiple times and only return the signature if their results are equal. However, it doesn't guarantee security in a case where adversary injects similar faults to multiple executions. Another efficient countermeasure technique to protect against faults is verification-after-sign but it cannot detect faults when injected into verification procedure such that it gives a pass to faulty signatures as valid signatures. While in presence of the proposed countermeasure, it thwarts the malicious attempt by detecting the data faults through redundant data slices. However, adversary can circumvent the countermeasure by injecting same faults in both the original data slices and their corresponding redundant data slices. If either an original data slice or its respective redundant data slice has an error, it is detected by taking their \texttt{XOR} and its expected value is non-zero. Whereas, in case of non-faulty operations, they will always produce a zero value. Since this fault-response generation step is not covered by IIR, they would be duplicated using conventional approaches to protect from instruction faults.  

We investigate the fault coverage of the proposed countermeasure using performance hotspots analysis. To perform this measurement on ARM Cortex-A9, we use Xilinx Vitis TCF Profiler. Results of profiling the protected Dilithium algorithm are shown in Table~\ref{tab:hotspotanalysis}. We observe that \textsc{butterflyCompute()} and \textsc{pointwisemultiplier()} bit-sliced functions are the largest contributors to CPU cycles with a CPU usage of $58.6\%$ and $20.4\%$ on ARM Cortex-A9 target platform. These two functions form the core of bit-sliced polynomial multiplications and protects the most time-consuming operations from data faults. Countermeasure also delivers a temporal fault coverage of about $78\%$, thereby, significantly decreasing the attack surface of Dilithium Algorithm. 

The proposed countermeasure doesn't provide protection against memory faults like Rowhammer \cite{2016rowhammerjs}. Typical countermeasure for such type of fault attacks is storing redundant copies of sensitive intermediate variables and verify its value before using it. Additionally, this countermeasure doesn't protect against instruction skipping faults since both original data slices and their corresponding redundant data slices will have the same effect. So, rejection sampling step implemented as an if-condition remains unguarded and if skipped, may leak secret-dependent signatures.

\renewcommand\arraystretch{1.1}
\begin{table*}[t]
\caption{\raggedright{Percentage of CPU cycles consumed across active functions and number of active functions calls during the combined execution of key generation and signature generation of Protected Dilithium on ARM Cortex-A9 CPU.}}
\label{tab:hotspotanalysis}
\centering
\vspace{0.1in}
\begin{tabular}{ccc}
    \toprule
    {\centering \textbf{Active Functions}} & {\centering \textbf{CPU Cycles (\%)}} & {\centering \textbf{No. of Calls}} \\ [9pt]
    \hline
    \textsc{butterflyCompute} & 58.6 & 16064 \\
    \hline
    \textsc{pointwisemultiplier} & 20.4 & 5360 \\
    \hline
    \textsc{pointwiseaccumulator} & 0.26 & 1024 \\
    \bottomrule
    \end{tabular}
\vspace{-0.1in}    
\end{table*}

\section{Experimental Results}
\label{section:results}
We perform performance \& footprint evaluation, overhead analysis and experimental fault attack resistance evaluation of the proposed countermeasure.
To inject faults, we use Electromagnetic Fault Injection (EMFI) setup as described in Section~\ref{section:fault_injection_setup}. 

\begin{figure}[t!]
    \centering
    \includegraphics[width=1.0\textwidth]{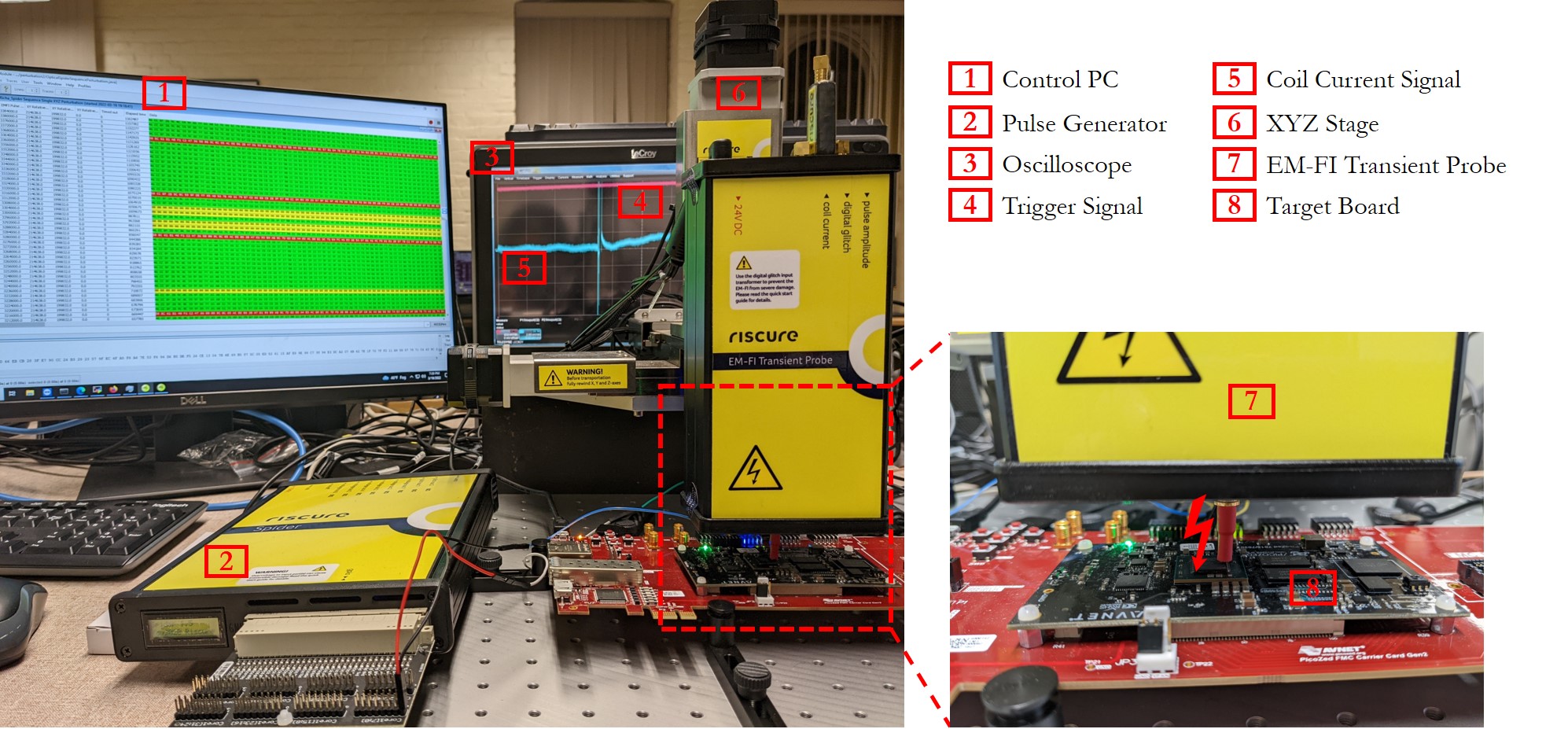}
    \caption{Overview photo of EMFI setup and close-up photo of the injection coil. The probe tip is positioned at approximately 0.4mm from the top of the processor package.}
    \label{fig:emfi_setup}
\end{figure} 

\subsection{Experimental Fault Injection Setup}
\label{section:fault_injection_setup}
This section describes our measurement setup for EM Pulse Injection. The setup consists of a hardware part, device under attack, a software part and experimental process. 

\subsubsection{Hardware Setup:} The EM fault injection bench is composed of a control PC, the targeted device, an automated XYZ stage, a pulse generator, an oscilloscope and an EMFI transient probe. We use commercially available FI hardware and software tools \cite{riscurefi} to build this setup. The target is placed on the XYZ stage as shown in the Figure \ref{fig:emfi_setup}. We use a classic EMFI probe tip made of a copper winding around a ferrite core. It has a flat-head tip of positive polarity and 1.5 mm diameter in order to cause disturbance in small part of the device. The probe is connected to the pulse generator and oscilloscope. Oscilloscope is used to measure the trigger and probe coil current signals. As soon as the probe receives a pulse from the pulse generator at its digital glitch input, it discharges the capacitor bank into the coil at the probe tip thereby creating a single EM pulse. The PC controls every part of the setup including configuring the injection parameters and capturing the results for analysis. The control software running on the PC synchronizes its operations with the other components of the setup through serial communication. 

\subsubsection{Device Under Attack:} 
The attack is realized on a Zynq-7000 SoC, XC7Z030SBG485 Flip-Chip Lidless BGA device having 19x19 mm package size. This SoC is embedded on Avnet PicoZed, our target board and it consists of a dual-core 32-bit ARM Cortex-A9 MPCore based processing system (PS) and Xilinx programmable logic (PL) on CMOS 28nm technology implementing the ARMv7-a ISA and runs by default at 667 MHz. The ARM processor has two separate 32 KB L1 caches for instruction and data, 256 KB on-chip memory and shares 512 KB L2 cache with NEON co-processor. The target board has 1 GB external DDR3 Memory and 128 MB QuadSPI Flash.  

\subsubsection{Software Setup:}
The test program is booted from external flash in PS Master Boot Mode and is executed from DDR3. We used Xilinx Vitis platform to program flash and for on-chip hardware debugging. We power the target during all the experiments at nominal voltage. A trigger is implemented using a general purpose IO pin of the target. The test program running on the target sets this trigger signal high just before the beginning of attack window and then sets the trigger signal low after the window ends. The test program used for all the experiments is open-source reference C implementation of Dilithium Round 3 version integrated with our countermeasure. This implementation was compiled with the arm-none-eabi-gcc-9.2.0 using compiler flags \texttt{-mcpu=cortex-a9 -mfpu=neon-vfpv3 -mfloat-abi=hard -O0}. For every fault injection, the test program sends a fault response and complete signature at the end of execution to the PC over UART for faults classification.

\subsubsection{Experimental Process:}
We configure four injection parameters in our EMFI setup for getting and increasing the probability of a successful fault injection. Since the full parameter search space is huge to be exhaustively covered, we rather focus on searching the optimal values for the most important parameters and keeping remaining parameters fixed:
\begin{itemize}
\item{\textbf{Spatial location -} The x-y position is defined as the 2D-position of injection probe relative to the reference points set on the top surface area of the chip. We keep the Z-position fixed at approximately 0.4mm from the top of the chip.}
\item{\textbf{Temporal location -}  The amount of time between the trigger is set high and the actual EM pulse injection. It is also referred to as \textbf{EMFI Pulse Delay}.}
\item{\textbf{Injection voltage -} The Intensity of EM pulse. It configures the maximum voltage over the coil which effects the current induced into the chip. It is also referred to as \textbf{EMFI Pulse Power} and this value is a percentage of the highest injection voltage of the probe, which is 475V.}
\item{\textbf{Pulse duration -} The amount of time the probe continuously supplies variable coil current. We fixed it to 500 kns.}
\end{itemize}

In this setup, the probe tip coil emits a single EM pulse for a fault injection. First, we gather golden signature response from the target running a fault-free execution for a fixed message and seed. After that, for every EM fault injection measurement, output response from the target is compared against the non-faulty output. The result of the experiment can be classified and grouped into different categories:  
\begin{itemize}
\item{\textbf{No Effect -} When the target output matches the expected response; no observable effect of fault is seen.}
\item{\textbf{Crash -} When the target results in error conditions which are handled by fault exceptions such as Data Abort, Pre-fetch Abort or an undefined instruction on ARM processors. For each of these exceptions, there is an exception handler set up with an infinite loop.}
\item{\textbf{Faults Not Detected -} When the output signature response is different from the expected response but fault response is not detected.}
\item{\textbf{Faults Detected -} When the output signature response is different from the expected response but fault response is detected.}
\end{itemize}

We do a software reset of the target after every measurement so that the result state of the previous measurement doesn't effect next measurement. After a reset, the target boots from flash and runs the test program automatically. In the crash cases, we do manual external reset of the target. Faulty signatures are those which pass the rejection checks in the signing procedure. In a real-world setting of the proposed countermeasure, when a fault is detected by our countermeasure, it would not output any faulty signature for secret key analysis by the attacker.

\subsection{Performance Analysis}
Table~\ref{tab:performance_result_arm} shows the performance impact of the proposed bit-slicing based IIR countermeasure on the key operations of Dilithium on ARM Cortex-A9 target platform. We report results for the NIST security level 2 of Dilithium (Dilithium2). These numbers are median cycle counts for 10,000 executions of Dilithium, including key generation and signing procedures only, as verification operates on public information. Note that hardware performance counters are used to measure cycles count. All the implementations for performance analysis are compiled with \texttt{-O3} optimization flag. Both the slowdowns are calculated by dividing by the corresponding unprotected Dilithium metric. For NTT/INTT, slowdown directly comes from the \textsc{butterflyCompute()} and \textsc{pointwisemultiplier()} bit-sliced functions since they involve a large number of data transfers as compared to compute cycles. Their contribution in runtime of Dilithium in terms of number of functions calls is shown in Table~\ref{tab:hotspotanalysis}. It is observed that Dilithium key generation and signing incurs high performance overhead since a significant portion of their runtime is dominated by NTT/INTTs, thereby, leading to large call counts of these bit-sliced functions. From the Table~\ref{tab:performance_result_arm}, it should be noticed that protected version have almost twice the slowdown compared to the bit-sliced version since the protection is based on dual spatial redundancy.

\renewcommand\arraystretch{1.1}
\begin{table*}[t]
\caption{Performance evaluation of bitslicing-based IIR countermeasure for key operations of Dilithium Algorithm on ARM Cortex-A9 @ 667MHz. The cycle counts are median for 10,000 executions. Unprotected Dilithium is the reference C implementation with no added countermeasure.}
\label{tab:performance_result_arm}
\centering
\vspace{0.1in}
\begin{tabular}{cccccc}
    \toprule
    \multirow{4}{*}{\parbox[t][][c]{2.0cm}{\centering \textbf{Dilithium Main Operations}}} & \multicolumn{3}{c}{\parbox[t][][c]{2.5cm}{\centering \textbf{Cycles Count}}} & \multicolumn{2}{c}{\parbox[t][][c]{2.5cm}{\centering \textbf{Slowdown ($\times$)}}} \\ [9pt]
    \cmidrule(lr){2-4}
    \cmidrule(lr){5-6}
    {} & \multirow{1.7}{*}{\parbox[t][][c]{1.91cm}{\centering \textbf{Unprotected}}} & \multirow{1.5}{*}{\parbox[t][][c]{1.5cm}{\centering \textbf{Bitsliced}}} & \multirow{1.5}{*}{\parbox[t][][c]{1.5cm}{\centering \textbf{Protected}}} & \multirow{1.5}{*}{\parbox[t][][c]{1.5cm}{\centering \textbf{Bitsliced}}} & \multirow{1.5}{*}{\parbox[t][][c]{1.5cm}{\centering \textbf{Protected}}} \\ [10pt]
    \hline
    Forward NTT & 28,928 & 925,824 & 1,827,584 & 32.0 & 63.2 \\
    \hline
    Inverse NTT & 30,848 & 928,128 & 1,833,152 & 30.1 & 59.4 \\
    \hline
    Point-wise Multiplication & 4,864 & 302,144 & 547,904 & 62.1 & 112.6 \\
    \hline
    Full Multiplication & 93,888 & 3,082,496 & 6,037,888 & 32.8 & 64.3 \\
    \hline
    Key Generation & 2,231,872 & 32,030,400 & 57,659,968 & 14.4 & 25.8 \\
    \hline
    Signature Generation & 7,150,272 & 191,078,144 & 351,530,304 & 26.7 & 49.2 \\
    \bottomrule
    \end{tabular}
\vspace{-0.1in}    
\end{table*}

\subsection{Footprint Analysis}
Footprint is calculated by measuring the compiled program size. Table \ref{tab:footprint_result} shows the footprint results for unprotected, bit-sliced and protected Dilithium implementations. All the implementations for footprint analysis are compiled with \texttt{-Os} optimization flag. Increased memory requirements in \texttt{.text} memory section for bit-sliced implementation comes from the \textsc{butterflyCompute()}, \textsc{pointwisemultiplier()} and \textsc{pointwiseaccumulator()} bit-sliced functions which occupy 77020 bytes, 82708 bytes and 3672 bytes respectively of code segment on ARM Cortex-A9 platform. The code size of a bit-sliced function is directly proportional to the cycles it takes to execute due to its linear code structure. This is evident by the protection cost of Dilithium implementation when protected with our bit-slicing based countermeasure mentioned in Table~\ref{tab:performance_result_arm}. While pre-computed twiddle and convolution factors used in bit-sliced polynomial multiplication computations are also part of \texttt{.text} section. Bit-sliced implementation also shows an overhead in \texttt{.bss} memory section due to uninitialized global intermediate array variables of polynomial size used in bit-sliced NTT/INTT operations. Table \ref{tab:footprint_result} shows that the footprint overhead of protected Dilithium implementation on target platform is 2.09 times.

\renewcommand\arraystretch{1.1}
\begin{table*}[t]
\caption{Footprint evaluation of bitslicing-based IIR countermeasure on ARM Cortex-A9 @ 667MHz target platform. Overhead is calculated by dividing the total code size by the corresponding unprotected implementation.}
\label{tab:footprint_result}
\centering
\vspace{0.1in}
\begin{tabular}{cccccc}
    \toprule
    \multirow{3}{*}{\parbox[t][][c]{2.0cm}{\centering \textbf{Dilithium Implementation}}} & \multicolumn{4}{c}{\parbox[t][][c]{4.0cm}{\centering \textbf{Code Size (bytes)}}} & \multirow{3}{*}{\parbox[t][][c]{1.5cm}{\centering \textbf{Overhead \\ ($\times$)}}} \\ [9pt]
    \cmidrule(lr){2-5}
    {} & \multirow{1.5}{*}{\parbox[t][][c]{1.5cm}{\centering \textbf{.text}}} & \multirow{1.5}{*}{\parbox[t][][c]{1.4cm}{\centering \textbf{.data}}} & \multirow{1.5}{*}{\parbox[t][][c]{1.5cm}{\centering \textbf{.bss}}} & \multirow{1.5}{*}{\parbox[t][][c]{1.5cm}{\centering \textbf{Total}}} & {} \\ [10pt]
    \hline
    Unprotected & 48,572 & 1,144 & 145,544 & 195,260 & - \\
    \cmidrule(lr){2-5}
    Bitsliced & 222,488 & 1,144 & 159,056 & 382,688 & 1.95 \\
    \cmidrule(lr){2-5}
    Protected & 232,064 & 1,144 & 176,464 & 409,672 & 2.09 \\
    \bottomrule
    \end{tabular}
\vspace{-0.1in}    
\end{table*}

\subsection{Overhead Analysis}
Table~\ref{tab:overhead_result} shows our analysis of different bit-sliced functions used in the Bit-sliced Dilithium implementation in terms of instruction breakdown and overhead of data movements. The overhead values reported are calculated as the number of \texttt{LDR} and \texttt{STR} instructions divided by the total number of instructions. We observe  the number of \texttt{ORR}, \texttt{EOR}, \texttt{AND}, \texttt{ADD} and \texttt{SUB} instructions involved during the execution of  \textsc{butterflyCompute()}, \textsc{pointwisemultiplier()} and \textsc{pointwiseaccumulator()} bit-sliced functions. Furthermore, we observe that moving data from memory to processor is expensive for these functions. The number and composition of logical bit-wise instructions originates directly from the Boolean gate-level netlist generated from the Verilog description. The overhead of \texttt{LDR} and \texttt{STR} instructions, however, is introduced by the compiler due to increased register pressure. There is register pressure because net width in the netlist, i.e., the number of active variables in bit-sliced code exceeds the number of registers available in the underlying hardware. The overhead related to register spilling into memory is about 48-62\% in terms of instruction count. This overhead explains the slower performance of the bit-sliced NTT, INTT, Full Multiplication and Matrix-Vector Multiplication operations of Dilithium. 

\renewcommand\arraystretch{1.1}
\begin{table*}[t]
\caption{Evaluation of different bit-sliced functions used in our Bit-sliced Dilithium Implementation on ARM Cortex-A9 @ 667MHz. Overhead is calculated as the number of ldr and str instructions as a percentage of the total}
\label{tab:overhead_result}
\centering
\begin{tabular}{ccccccccc}
     \toprule
     {\parbox[t][][c]{2.5cm}{\centering \textbf{Bit-sliced \\ function}}} & \multicolumn{7}{c}{\parbox[t][][c]{2.5cm}{\centering \textbf{Instruction Mix}}} & {\parbox[t][][c]{1.5cm} {\centering \textbf{Overhead (\%)}}} \\[5pt]
     \cline{2-8}
     {} & ORR & EOR & AND & ADD & SUB & LDR & STR & {}  \\
     \hline
     \textsc{butterflyCompute} & 1830 & 2957 & 3994 & 2 & 2 & 5210 & 3130 & 48.70 \\
     \hline
     \textsc{pointwisemultiplier} & 1931 & 3270 & 4331 & 2 & 2 & 5618 & 3295 & 48.31 \\
     \hline
     \textsc{pointwiseaccumulator} & 51 & 97 & 120 & 13 & 1 & 277 & 194 & 62.54 \\
     \bottomrule
 \end{tabular}
\end{table*}

\subsection{Fault Countermeasure Evaluation}
In this section, our goal is to maximize the chances of obtaining a successful fault which gets detected by the proposed countermeasure. For this, determination of injection parameters is an essential step which consists of finding the spatial location, temporal location and injection voltage. We demonstrate the effectiveness of the proposed countermeasure by performing fault injection on protected NTT of \textsc{$s_1$}, our point of interest in the test program and it is timed using a GPIO-based trigger as explained in Section \ref{section:fault_injection_setup}. Protected NTT is divided into three phases: transpose, compute and reverse-transpose. As we know that redundant computations take place in the compute phase, therefore, we set compute phase as our attack window and it takes roughly 10M ns.

We first performed full chip scan with a 20 x 20 grid resolution and EMFI Pulse Delay and EMFI Pulse Power were randomly set to values between 0 ns \& 10M ns and 70\% \& 100\% respectively for each measurement. We observed through this initial trial that successful faults are obtained in the right upper quadrant of the chip which also correlates to an area containing target Processing System (PS). Also, there were no successful faults when EMFI Pulse Power is less than 80\%. With this partial identification of successful faults distribution, we perform three experiments to fine-tune our injection parameters in a top-down approach:

\subsubsection{Experiment 1: Randomly-chosen EMFI Pulse Power and EMFI Pulse Delay from an estimated range with a step-wise right upper quadrant chip scan}
\begin{figure}[t!]
    \centering
    \includegraphics[width=1.0\columnwidth]{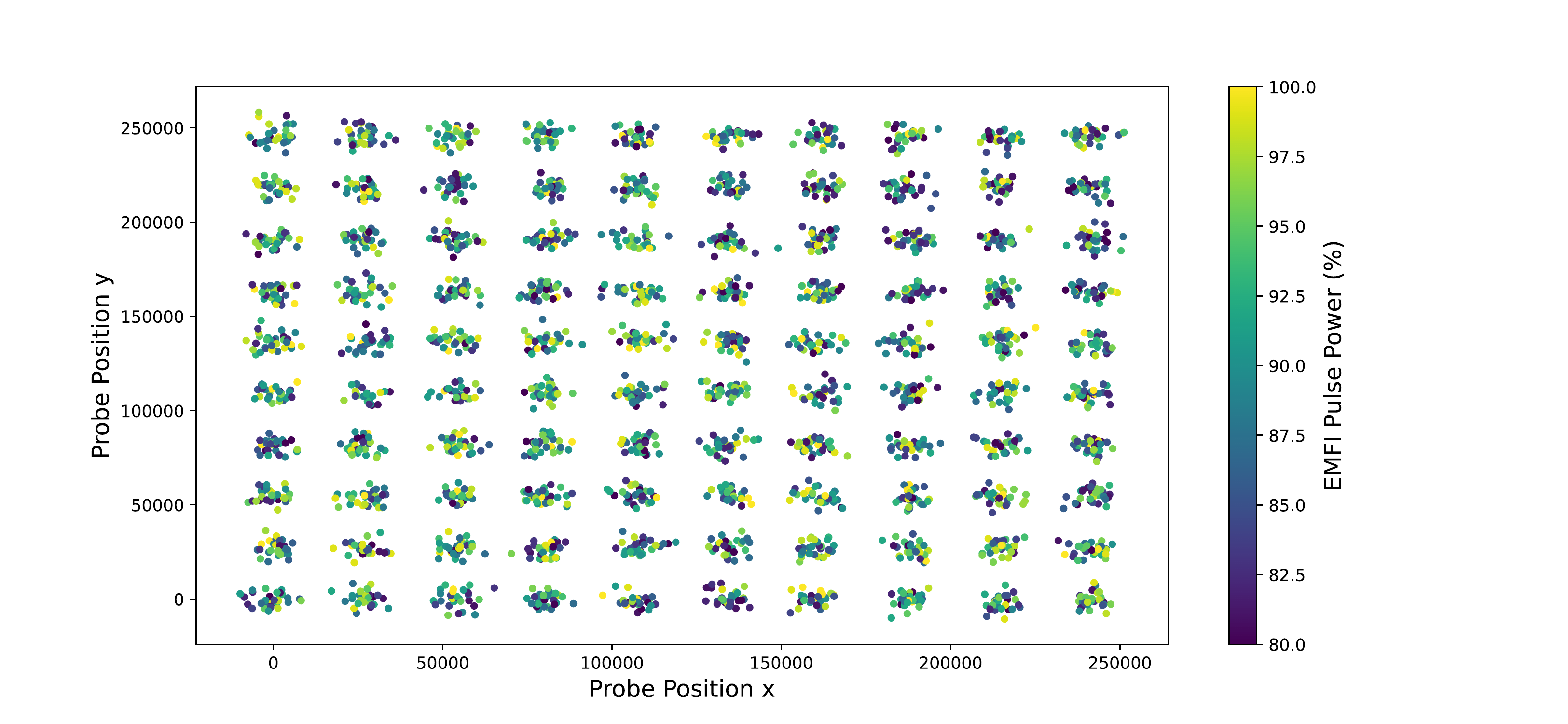}
    \caption{Experiment 1: Probe positions over the chip leading to "No Effect" fault cases. Note that it is a jittered scatter plot to prevent overlapping dots at the same position.}
    \label{fig:scatterheatmap_greenfaults}
\end{figure}

\begin{figure*}[t!]
    \centering
    \begin{subfigure}[t]{0.49\textwidth}
        \centering
        \includegraphics[width=\textwidth]{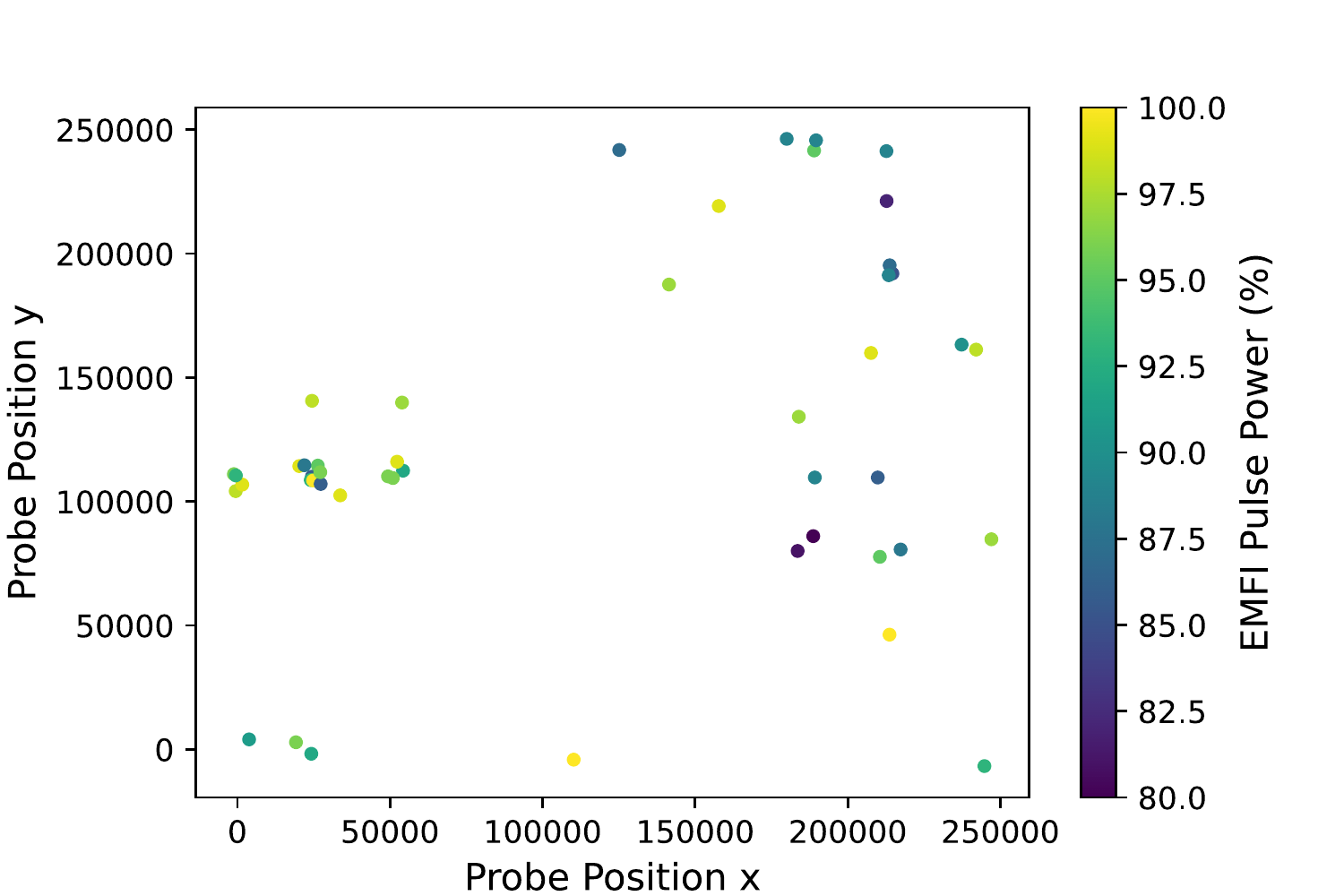}
        \caption{Grouped "Faults Detected", "Faults Not Detected" and "Crash" cases}
        \label{fig:scatterheatmap_red_white_yellowfaults}
    \end{subfigure} %
    ~ 
    \begin{subfigure}[t]{0.49\textwidth}
        \centering
        \includegraphics[width=\textwidth]{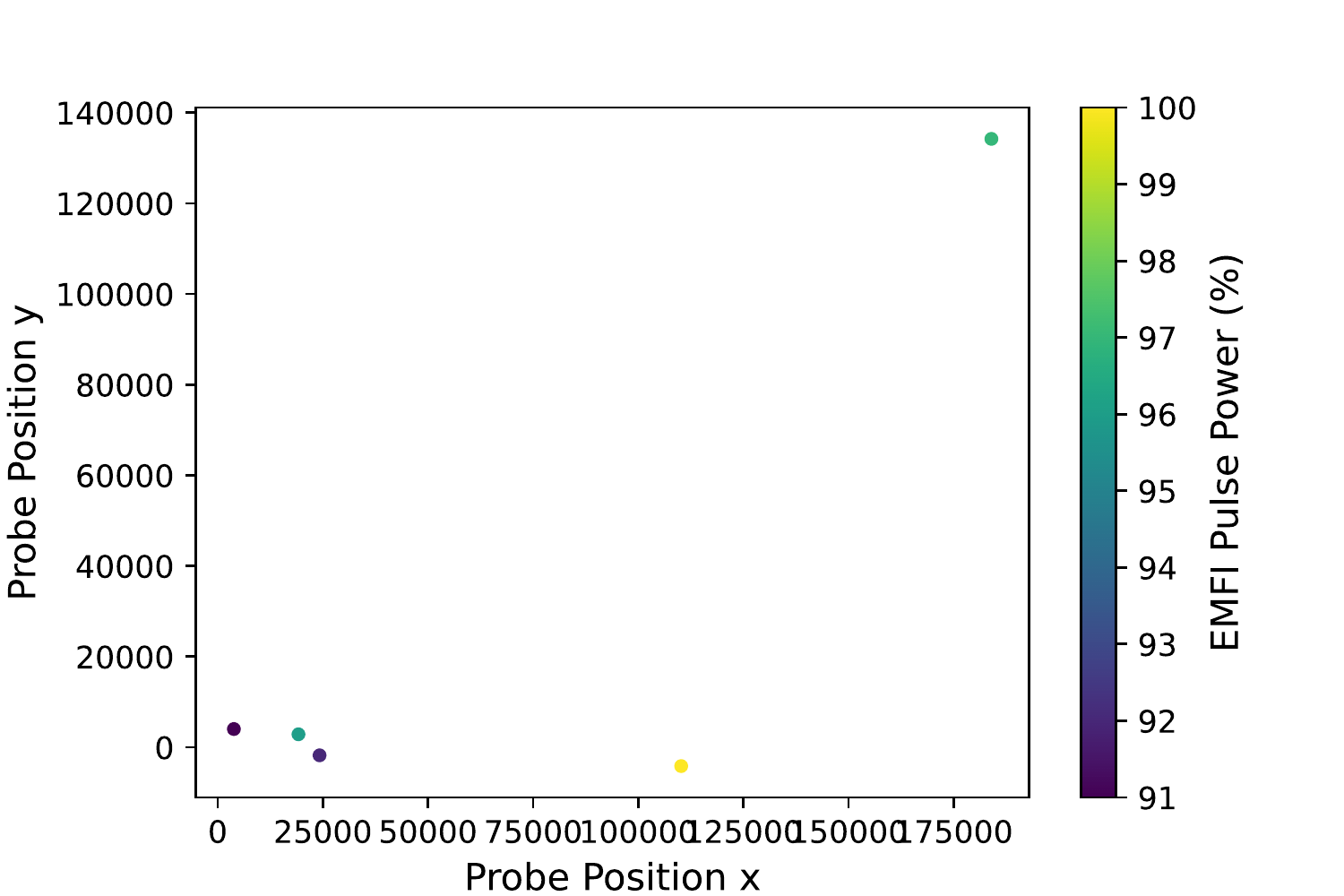}
        \caption{"Crash" cases}
        \label{fig:scatterheatmap_yellowfaults}
    \end{subfigure} 
    \hfill
    \begin{subfigure}[t]{0.49\textwidth}
        \centering
        \includegraphics[width=\textwidth]{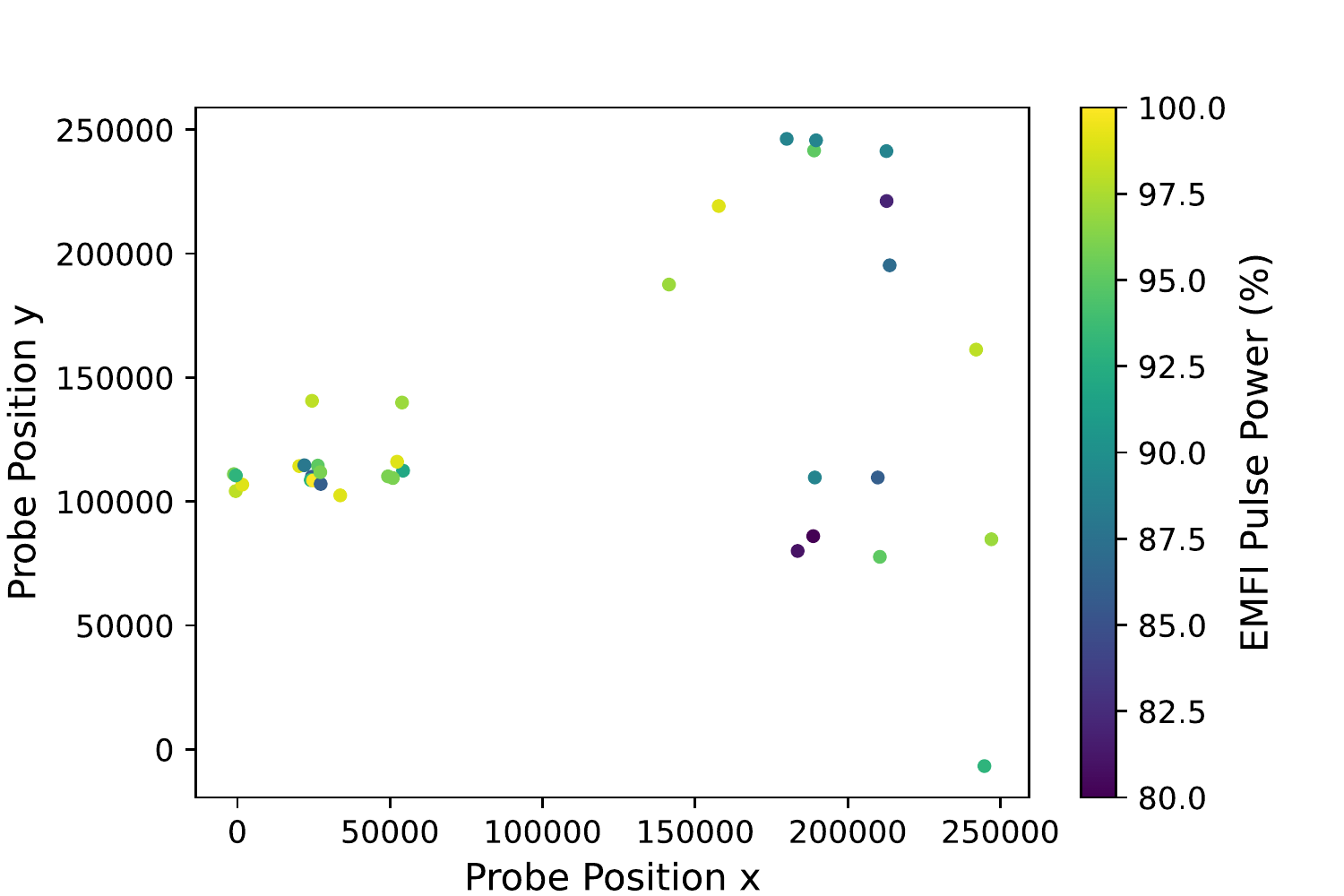}
        \caption{"Faults Not Detected" cases}
        \label{fig:scatterheatmap_whitefaults}
    \end{subfigure} %
    ~ 
    \begin{subfigure}[t]{0.49\textwidth}
        \centering
        \includegraphics[width=\textwidth]{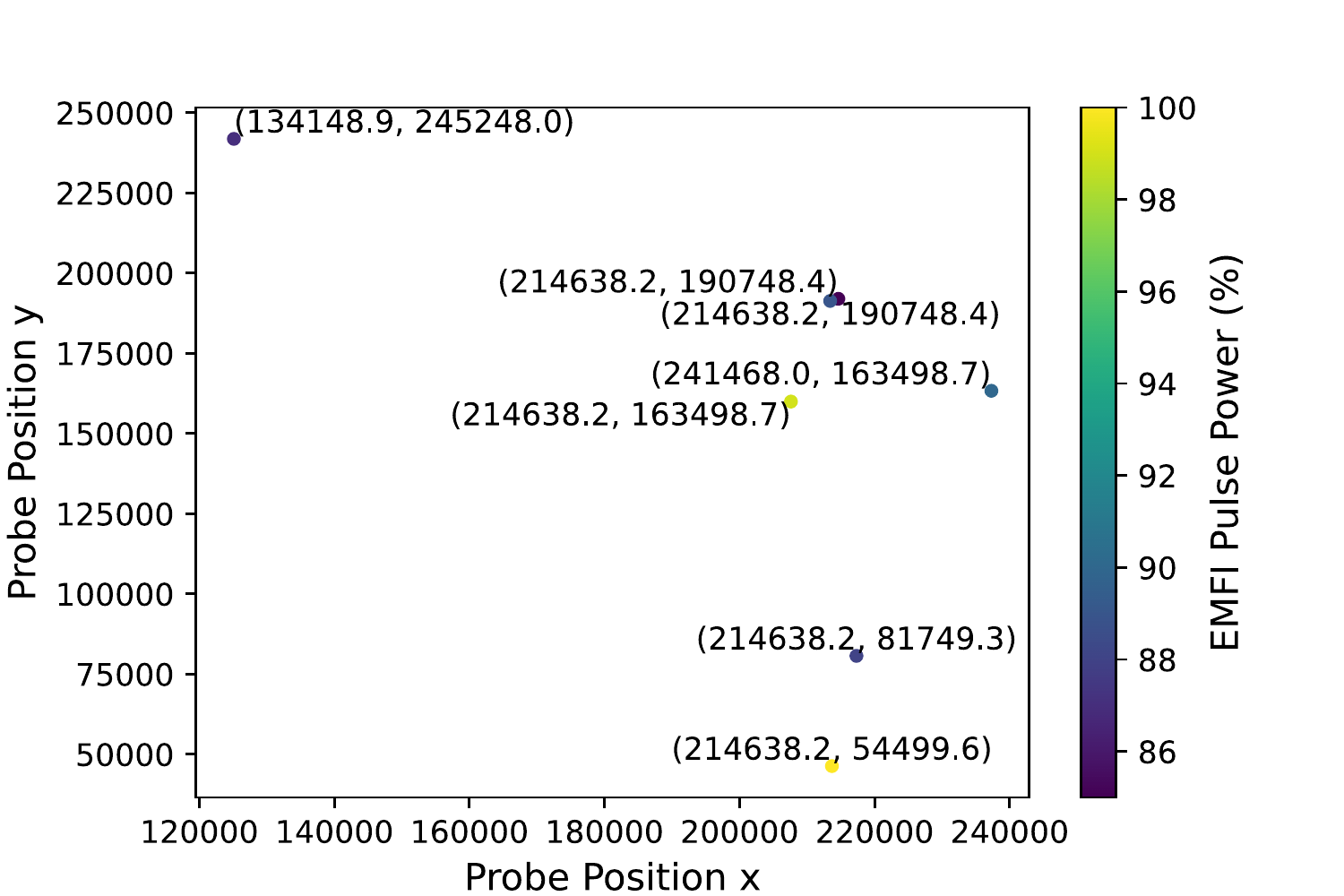}
        \caption{"Faults Detected" cases}
        \label{fig:scatterheatmap_redfaults}
    \end{subfigure} %
    ~
        \caption{Experiment 1: Probe positions over the chip leading to fault cases which change the test program's intended behavior. Note that it is a jittered scatter plot to prevent overlapping dots at the same position.}
\end{figure*}

\mbox{}
\label{par:exp1}
\newline
To determine the best probe x-y position within right upper quadrant of the chip, we performed scan in a 10 x 10 measurement grid and injected 30 faults per probe position, this led to 3000 measurements in total. The configuration of injection parameters for this experiment are summarized in the Table \ref{tab:exp1_param}. Table \ref{tab:exp1_perfaults} lists the outcome for this surface area exploration while injecting into the test program in terms of percentage faults in each category. Areas where no effect of fault injection is seen are shown in the Figure \ref{fig:scatterheatmap_greenfaults} while positions which are sensitive to fault injection where faults are either resulting in a crash, faults being detected and faults not detected are shown in the Figure \ref{fig:scatterheatmap_red_white_yellowfaults}. We can also observe the spatial occurrence of individual fault categories where an effect of fault injection is visible in the following Figures \ref{fig:scatterheatmap_yellowfaults}, \ref{fig:scatterheatmap_whitefaults} and \ref{fig:scatterheatmap_redfaults}.

\begin{figure}[t!]
    \centering
    \includegraphics[width=1.0\columnwidth]{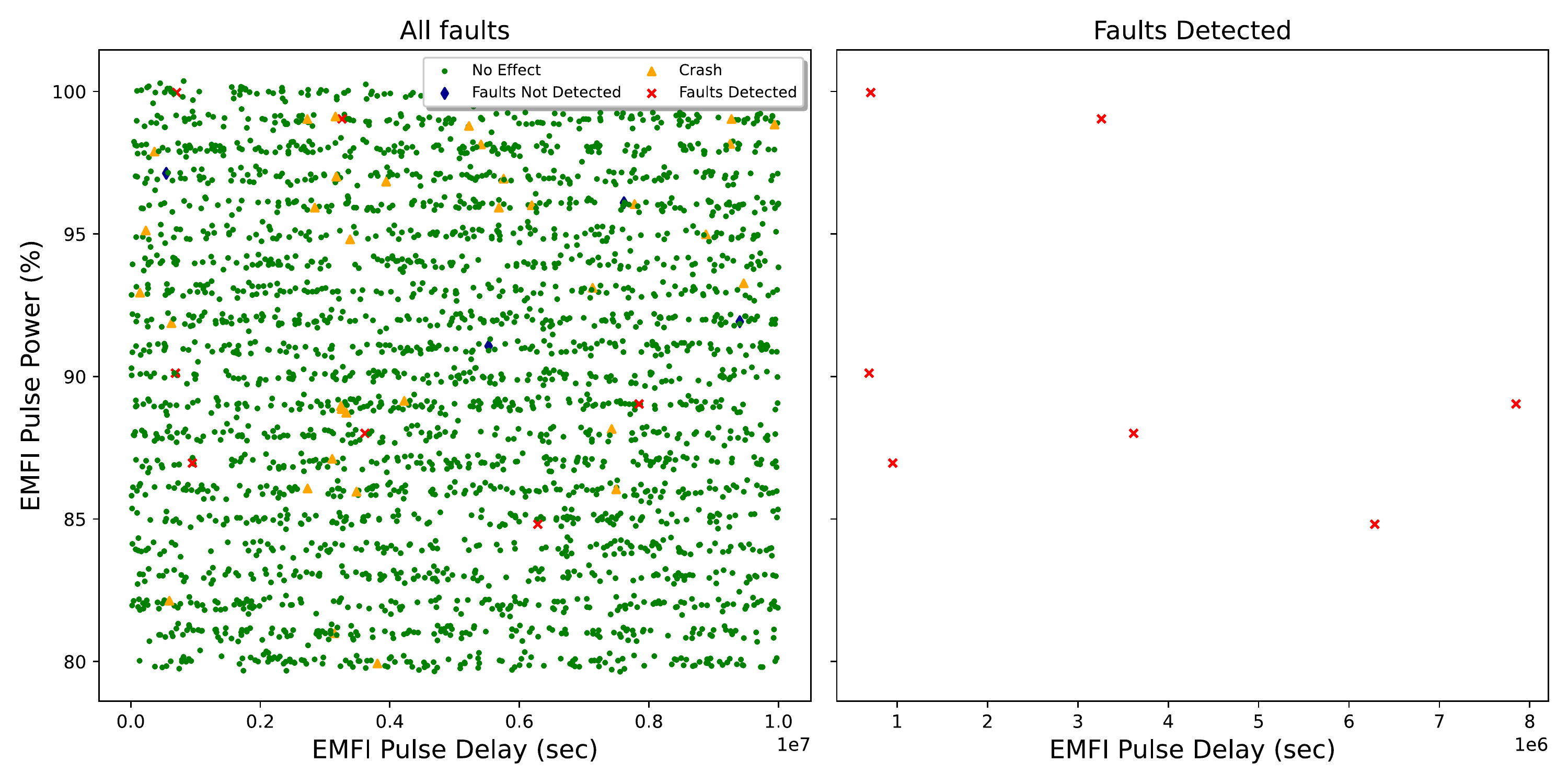}
    \caption{Experiment 1: Relation between EMFI Pulse Delay and EMFI Pulse Power for faults of different categories and only "Faults Detected" category is shown in the left and right plot respectively.}
    \label{fig:EMPIvsEMPD_random_all_redfaults_exp1}
\end{figure}

We can identify the best probe x-position as 214638 since 5 out 7 faults detected occur at the same x-position while best y-position is 199832 by taking the average of y co-ordinates of faults detected that fall in the top-half area. This means the probe x-y position can be fixed to decrease the spatial location parameter search space, significantly increasing the faults detection rate. We can conclude from Figure \ref{fig:EMPIvsEMPD_random_all_redfaults_exp1} that there are more faults detected when injection is performed between 0 to 5M ns duration of attack window and EMFI Pulse Power is less than 90\%. At higher pulse intensities, more crash behavior is observed while lower pulse intensities does not cause enough voltage variations in the device to induce computational faults.

\renewcommand\arraystretch{1.1}
\begin{table*}[t]
\caption{Experiment 1: Injection parameters settings}
\label{tab:exp1_param}
\centering
\begin{tabular}{|c|c|}
     \hline
     \textbf{Injection Parameter} & \textbf{Value} \\
     \hline
     x-y Probe Position & Right Upper Quadrant Scan  \\
     \hline
     EMFI Pulse Delay & Random between 0 ns and 10M ns \\
     \hline
     EMFI Pulse Power & Random between 80\% and 100\% \\
     \hline
 \end{tabular}
\end{table*}

\renewcommand\arraystretch{1.1}
\begin{table*}[t]
\caption{Experiment 1: Percentage occurrence of different fault categories}
\label{tab:exp1_perfaults}
\centering
\begin{tabular}{ccc}
     \toprule
     {\parbox[t][][c]{2.1cm}{\centering \textbf{Classification}}} & {\parbox[t][][c]{2.1cm}{\centering \textbf{Amount}}} & {\parbox[t][][c]{2.3cm}{\centering \textbf{Percentage (\%)}}} \\ [10pt]
     \hline
     No Effect & 2953 & 98.4333 \\
     \hline
     Crash & 35 & 1.16667 \\
     \hline
     Faults Not Detected & 5 & 0.16666 \\
     \hline
     Faults Detected & 7 & 0.23333 \\
     \bottomrule
 \end{tabular}
\end{table*}

\subsubsection{Experiment 2: Randomly-chosen EMFI Pulse Power and EMFI Pulse Delay within an identified range with a fixed x-y probe position}
\begin{figure}[t!]
    \centering
    \includegraphics[width=1.0\columnwidth]{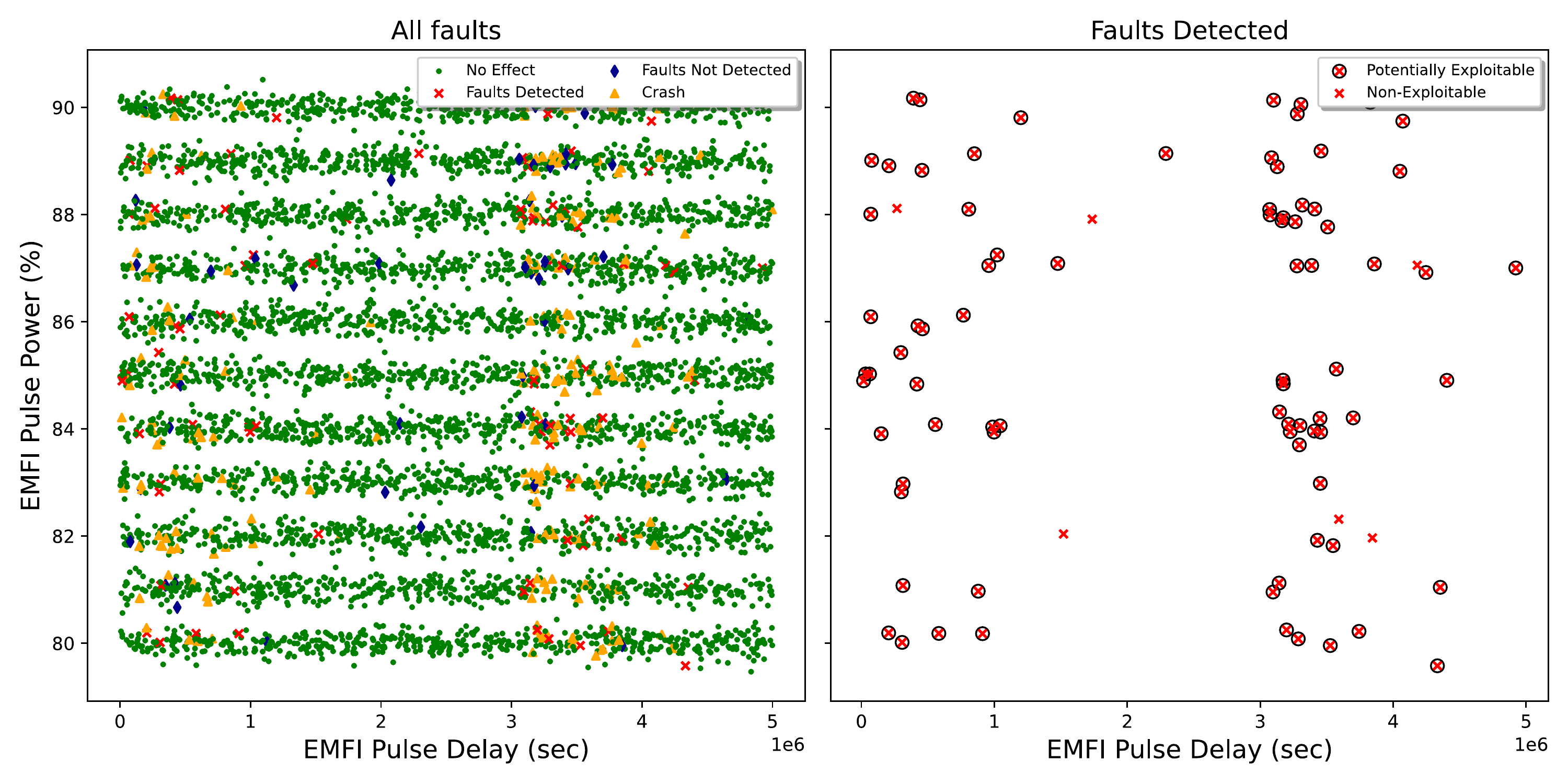}
    \caption{Experiment 2: Relation between EMFI Pulse Delay and EMFI Pulse Power for faults of different categories and only "Faults Detected" category is shown in the left and right plot respectively.}
    \label{fig:EMPIvsEMPD_random_all_redfaults}
\end{figure}

\mbox{}
\label{par:exp2}
\newline
This experiment is performed at a fixed spatial location determined through Experiment \ref{par:exp1}. Also, broad EMFI delay and intensity spectrum has been reduced from previous experiment and those settings are summarized in the Table \ref{tab:exp2_param}. Using these parameters, we investigate the relationship between the EMFI Pulse Delay and EMFI Pulse Power to find an optimal parameter range in order to enhance the probability of detected successful faults. We perform total 6000 measurements in 21 hours and the results of this experiment are shown in Figure \ref{fig:EMPIvsEMPD_random_all_redfaults} and Table \ref{tab:exp2_perfaults}.

The plot shows that a lot of faults detected are spread out in a time window of 3M ns to 3.8M ns after the trigger. Also, we factorize the faults into two categories: potentially exploitable and non-exploitable. Potentially exploitable faults are usable for an attacker if the test program executes completely but computes a non-zero faulty signature whereas zero faulty signatures and correct signatures are non-exploitable. Crashes are faults aborting the normal execution of test program hence are not usable for an attack. Note that we get few cases where faults get detected but the signature response is as expected. These are also counted as non-exploitable faults being detected. We are able to find an easily glitchable area in the plot when EMFI Pulse Power is between 83\% and 90\% where we can spot clustering of fault detected red crosses. We can observe from the plot that there is a high correlation between the sensitive area for crashes and the sensitive area for potentially exploitable faults. To conclude, this experiment shows that the percentage of potentially exploitable faults that are data faults is 63.77\%.

\renewcommand\arraystretch{1.1}
\begin{table*}[t]
\caption{Experiment 2: Injection parameters settings}
\label{tab:exp2_param}
\centering
\begin{tabular}{|c|c|}
     \hline
     \textbf{Injection Parameter} & \textbf{Value} \\
     \hline
     x-y Probe Position & Fixed at (214638, 199832) \\
     \hline
     EMFI Pulse Delay & Random between 0 ns and 5M ns \\
     \hline
     EMFI Pulse Power & Random between 80\% and 90\% \\
     \hline
 \end{tabular}
\end{table*}

\renewcommand\arraystretch{1.1}
\begin{table*}[t]
\caption{Experiment 2: Percentage occurrence of different fault categories and their breakup into potentially exploitable faults and non-exploitable faults}
\label{tab:exp2_perfaults}
\centering
\begin{tabular}{ccccc}
     \toprule
     {\parbox[t][][c]{2.1cm}{\centering \textbf{Classification}}} & {\parbox[t][][c]{2.1cm}{\centering \textbf{Amount}}} & {\parbox[t][][c]{2.1cm}{\centering \textbf{Percentage (\%)}}} & {\parbox[t][][c]{2.5cm}{\centering \textbf{Potentially Exploitable}}} &
     {\parbox[t][][c]{2.5cm}{\centering \textbf{Non-Exploitable}}} \\ [15pt]
     \hline
     No Effect & 5635 & 93.9157 & 0 & 5635 \\
     \hline
     Crash & 229 & 3.81667 & 0 & 229 \\
     \hline
     Faults Not Detected & 49 & 0.816667 & 46 & 3 \\
     \hline
     Faults Detected & 87 & 1.45 & 81 & 6 \\
     \bottomrule
 \end{tabular}
\end{table*}

\subsubsection{Experiment 3: Parametric-sweep of EMFI Pulse Power and EMFI Pulse Delay with a fixed x-y probe position}
\begin{figure}[t!]
    \centering
    \includegraphics[width=1.0\columnwidth]{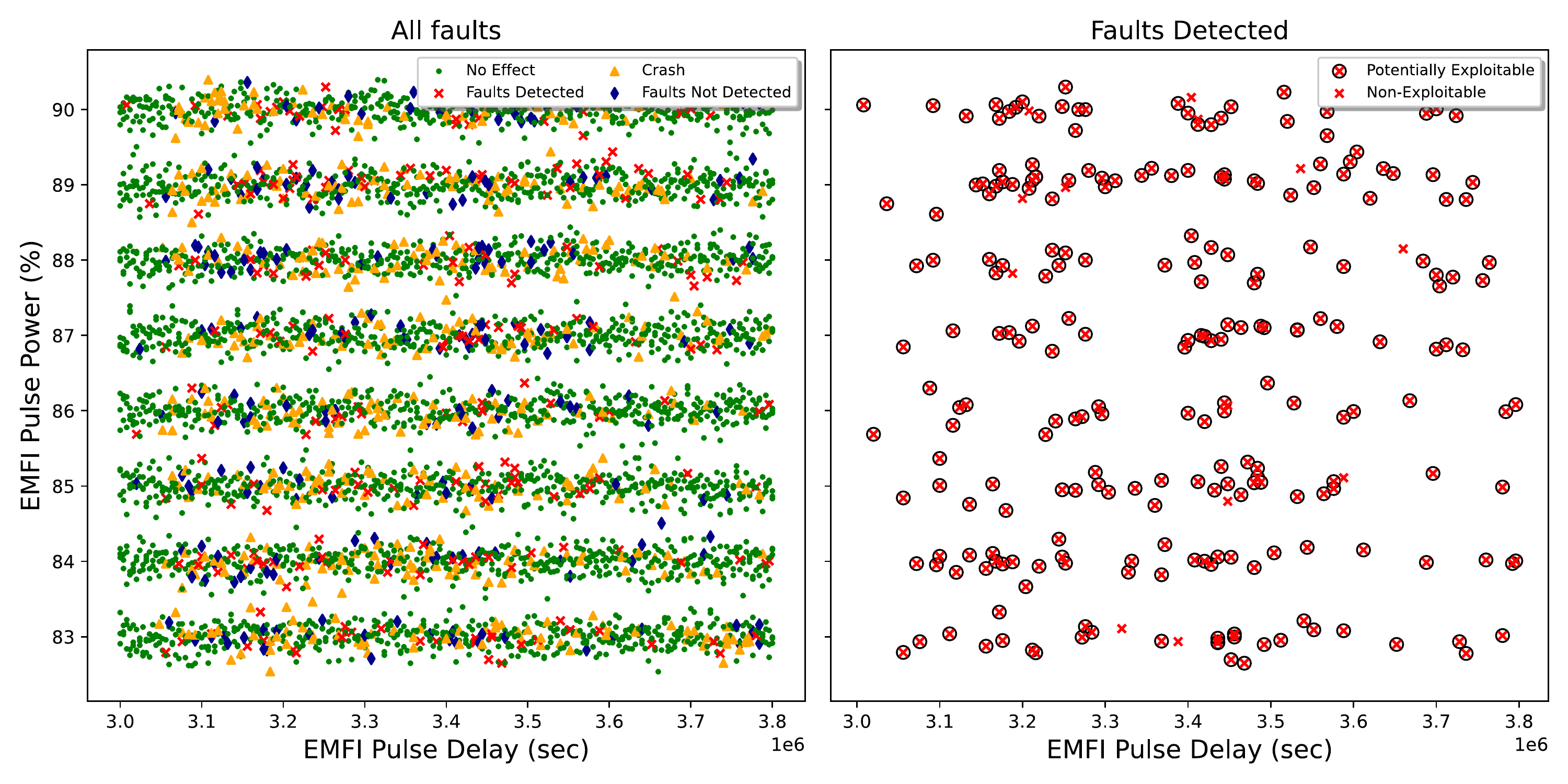}
    \caption{Experiment 3: Relation between EMFI Pulse Delay and EMFI Pulse Power for faults of different categories and only "Faults Detected" category is shown in the left and right plot respectively.}
    \label{fig:EMPIvsEMPD_range_all_redfaults}
\end{figure}

\mbox{}
\newline
In this experiment, we perform parameter sweep between the narrowed-down ranges identified through Experiment \ref{par:exp2}. Using the parameters given in Table \ref{tab:exp3_param}, we perform 3 measurements per step that leads to total 4824 measurements completed in 16 hours. The results of this experiment are shown in the Figure \ref{fig:EMPIvsEMPD_range_all_redfaults} and Table \ref{tab:exp3_perfaults} further shows the distribution of fault response.

It is observed that the number of potentially exploitable faults has increased from 127 in Experiment \ref{par:exp2} to 382 in this experiment. This result further confirms that the spatial parameter search process led to a precise location on the chip where we are able to achieve large number of attacker usable faults. The plot shows the same observation as in previous experiments that the attacker usable faults occur at similar locations or locations closer to crashes. In conclusion, our approach to injection parameters exploration is able to find reliable and highly repeatable potentially exploitable faults which affect the datapath are 61.7\% while the remaining are control flow faults or memory faults.

\renewcommand\arraystretch{1.1}
\begin{table*}[t]
\caption{Experiment 3: Injection parameters settings}
\label{tab:exp3_param}
\centering
\begin{tabular}{|c|c|}
     \hline
     \textbf{Injection Parameter} & \textbf{Value} \\
     \hline
     x-y Probe Position & Fixed at (214638, 199832) \\
     \hline
     EMFI Pulse Delay & Sweep between 3M ns and 3.8M ns by steps of 4000 ns \\
     \hline
     EMFI Pulse Power & Sweep between 83\% and 90\% by steps of 1\% \\
     \hline
 \end{tabular}
\end{table*}

\renewcommand\arraystretch{1.1}
\begin{table*}[t]
\caption{Experiment 3: Percentage occurrence of different fault categories and their breakup into potentially exploitable faults and non-exploitable faults}
\label{tab:exp3_perfaults}
\centering
\begin{tabular}{ccccc}
     \toprule
     {\parbox[t][][c]{2.1cm}{\centering \textbf{Classification}}} & {\parbox[t][][c]{2.1cm}{\centering \textbf{Amount}}} & {\parbox[t][][c]{2.1cm}{\centering \textbf{Percentage (\%)}}} & {\parbox[t][][c]{2.5cm}{\centering \textbf{Potentially Exploitable}}} &
     {\parbox[t][][c]{2.5cm}{\centering \textbf{Non-Exploitable}}} \\ [15pt]
     \hline
     No Effect & 3762 & 77.9851 & 0 & 3762 \\
     \hline
     Crash & 583 & 12.0854 & 0 & 583 \\
     \hline
     Faults Not Detected & 230 & 4.76783 & 146 & 84 \\
     \hline
     Faults Detected & 249 & 5.16169 & 236 & 13 \\
     \bottomrule
 \end{tabular}
\end{table*}

\section{Conclusion}
\label{section:conclusion}
We conclude that hardening the polynomial multiplication operations in Dilithium with the proposed countermeasure significantly reduced the probability of a successful fault injection attack and increased the time required to gain enough faulty signatures to mount a complete attack at the cost of increased computational complexity. FI attack surface for Dilithium is so large that any part of code may become a potential FI attack vector. Our countermeasure reduces the attack surface by protecting polynomial multiplication operations, second largest contributor to the run-time of the algorithm after hash operations. The proposed countermeasure estimates the percentage of potentially exploitable faults which affect the datapath as 62\%. However, some exploitable faults are not detected because time redundancy is not implemented in the proposed countermeasure to detect control flow faults. Further, characterization of fault effects is required to better understand the behavior of undetected potentially exploitable faults at software level. We plan to extend our bit-sliced NTT design with temporal IIR to detect faults in the control flow. Also, we intend to investigate the fault attack resistance of other finalist lattice-based schemes with our generic countermeasure.


\bibliographystyle{ACM-Reference-Format}
\bibliography{sample-acmsmall}


\end{document}